\documentclass[10pt]{article}

\usepackage{epsfig}
\usepackage{amsmath}
\usepackage{amsfonts}
\usepackage{amssymb}
\usepackage{multirow}
\usepackage{hyperref}

 \def    \ptj           {\mbox{$p_{T j}$}}
 \def    \etaj           {\mbox{$\eta_{ j}$}}
 \def    \ptl           {\mbox{$p_{T l}$}}
 \def    \etal           {\mbox{$\eta_{ l}$}}
 \def    \gev            {\mbox{$\mathrm{GeV}$}}

\newcommand{\etmiss}{ \not \hskip -4pt E_T}
\newcommand{\ptmiss}{ \not \hskip -4pt p_T}

\begin{document}

\setcounter{page}{0}
\thispagestyle{empty}


\vskip 8pt

\begin{center}
{\bf \Large {
New W-prime signals at the LHC
}}
\end{center}

\vskip 10pt

\begin{center}
{\large Natascia Vignaroli}
\end{center}

\vskip 20pt

\begin{center}

\centerline{Department of Physics and Astronomy, Michigan State University, East Lansing, MI 48824, USA}
\vskip 3pt
\vskip .3cm
\end{center}

\vskip 50pt

\begin{abstract}
We study the $W^{'}$ phenomenology in composite Higgs/ warped extra dimensional models focusing on the effect of fermionic resonances at $\sim$1 TeV. 
After deriving the existing bounds from the current LHC-8 analyses, we highlight the most promising signatures for $W^{'}$ discovery at the 14 TeV LHC. We find in particular very promising the study of $W^{'}$ decay modes into vector-like top partners, specifically the decay into a doublet of custodian heavy fermions, $T_{5/3} T_{2/3}$, and the decay into a heavy fermion plus a Standard Model quark. We perform a detailed parton level analysis of the channel $W^{'}\to T_{5/3} T_{2/3}$ in the same-sign dilepton final state, finding that it is a very promising signature to test the region at high $W^{'}$ mass, $m_{W^{'}}\gtrsim 2$ TeV, and of the $W^{'}\to Tb$ mode, that is one of the best channels to test the intermediate $W^{'}$ mass region and that, already with the LHC-8 data, could extend the present exclusion bounds. 
\end{abstract}

\vskip 13pt

\newpage

\tableofcontents


\section{Introduction}
\label{sec:introduction}

We study in this work the phenomenology of a $W^{'}$ boson from a beyond-the-SM $SU(2)_L$ triplet in the context of composite Higgs/ warped extra dimensional theories with custodial symmetry protection \cite{Agashe:2004rs, Agashe:2003zs}. In this scenario the $\sim$125 GeV particle discovered at the LHC \cite{Aad:2012tfa, Chatrchyan:2012ufa} can be interpreted as the pseudo-Goldstone Boson associated with a global invariance of a new strong dynamics, composite at the TeV scale. The Goldstone nature of the Higgs in addition to its compositeness can naturally account for a $\sim$125 GeV mass if the divergences to the Higgs mass are cut-off by top-partner vectro-like quarks below $\sim$1 TeV (see for example \cite{Panico:2012uw} and references therein). On the other hand, electro-weak-precision-tests, and in particular the bounds on the $S$ parameter, indicate a mass above $\sim$2 TeV for the new spin-1 resonances from the strong electroweak sector \cite{Ciuchini:2013pca, Grojean:2013qca}. These two indications together point toward a scenario where the $W^{'}$ is above the threshold for the decay into a top partner or even into a pair of vector-like quarks. We will thus analyze the $W^{'}$ phenomenology considering the presence of top-partners at $\sim$1 TeV and focusing, in particular, on the  $W^{'}$ decays into these vector-like quarks.\\
After deriving the bounds on $W^{'}$ mass and couplings from LHC-8 searches, we will perform detailed analyses of the $W^{'}$ decay  
into a top-prime plus a bottom quark, $Tb$, in the semileptonic channel and of the $W^{'}$ decay into a pair of custodians, $T_{5/3}T_{2/3}$, in the same-sign dilepton final state. The custodians, which are the lightest $SU(2)_L$ doublet top-partners, are a general prediction of composite/RS models with custodial symmetry \cite{Contino:2006qr}. The $W^{'}$ to custodians channel can thus represent a smoking-gun signature for this wide class of BSM theories.\\
The `new' $W^{'}$ to top-partners signals have been so far overlooked by the searches for W-primes at the LHC, which have been focused on `standard' decay modes, $W^{'}\to l\nu, jj, WZ$, motivated mainly by the sequential-SM hypothesis \cite{Altarelli:1989ff}, or on the $tb$ channel, considering composite/RS scenarios with a large top degree of compositeness/ top profile peaked near the TeV brane \cite{Agashe:2008jb}. We will prove in this work the high discovery potential of the new $W^{'}$ signals, that can widely test the composite Higgs hypothesis up to very heavy  ($\sim 3.5$ TeV) W-primes at the 14 TeV LHC.  \\  

The paper is organized as follows: we will briefly introduce the model and discuss the $W^{'}$ phenomenology, highlighting the most promising signatures for the $W^{'}$ discovery at the LHC, in sec. \ref{sec:model}; in Sec. \ref{sec:limits} we will present the bounds on $W^{'}$ mass and couplings derived from LHC-8 analyses; we will perform detailed parton-level analyses of the heavy-light $W^{'}\to Tb$ channel and of the custodian channel, showing the corresponding discovery/exclusion potentials, in Sec. \ref{sec:heavylight} and \ref{sec:custodian}; we will conclude in Sec. \ref{sec:conclusion}. In the App. \ref{sec:app} we will briefly describe the phenomenology of a $W^{'}_R$ from a $SU(2)_R$ BSM triplet and its mixing with the $W^{'}$.

\section{$W^{'}$ phenomenology}
\label{sec:model}

We discuss in this section the phenomenology of a W-prime vector boson in pseudo-Goldstone composite Higgs models with custodial symmetry \cite{Agashe:2004rs}, which is dual to that of a Kaluza-Klein W in warped extra-dimensional theories with a custodial symmetry in the bulk \cite{Agashe:2003zs}. \\
The relevant phenomenology can be captured by a simple two-site description \cite{Contino:2006nn} where the SM particles and the new heavy vector and fermionic resonances result from the mixing of a weakly-coupled sector of elementary particles with a strongly-coupled sector, made up of states which become composite at the TeV scale and which includes the Higgs. This determines a scenario of partial compositeness of the SM particles, where the heavier states, like the top, can have a large degree of compositeness. The corresponding RS dual picture is that of a top localized near the TeV brane. \\
Motivated by minimal $SO(5)/SO(4)$ theories, 
we consider a $SU(2)_L \times SU(2)_R\times U(1)_X$ global invariance of the strong dynamics. The $SU(2)_L \times U(1)_Y$ elementary bosons gauge the corresponding global invariance with $Y=T^3_R+X$. The strong sector includes a composite spin-1 resonance which transforms as $(\bf 3, \bf 1)_{0}$ under $SU(2)_L \times SU(2)_R\times U(1)_X$ and which mixes, analogously to $\rho$-photon mixing, with the elementary $SU(2)_L$ triplet. The mixing is diagonalized by a field rotation from the elementary-composite basis to the basis of mass eigenstates, which include the SM $W$ and the $W^{'}$ as superpositions of elementary and composite states. The $W^{'}$ will be given by
\begin{equation}\label{eq:rotation}
W^{'}=\cos\theta_2 W^{'}_{com} +\sin\theta_2 W_{el} \ ,
\end{equation}   
and the SM $W$ by the orthogonal combination.
The rotation is determined by

\begin{equation}\label{eq:ct2}
\cot\theta_2=\frac{g^{*}_2}{g^{el}_{2}} \qquad g_2=g^{el}_{2}\cos\theta_2=g^{*}_{2}\sin\theta_2 \ ,
\end{equation}   
\noindent
where $g_2=e/\sin\theta_W$ is the SM gauge coupling and $g^{el}_{2}$ and $g^{*}_2$ are respectively the coupling of the $SU(2)_L$ in the elementary sector and in the composite sector. The $\theta_2$ angle controls the strength of $W^{'}$ interactions with the SM particles and with the strong sector; the $W^{'}$ couplings to composite modes are proportional to $\cot\theta_2$, those to elementary modes to $\tan\theta_2$. Larger $\cot\theta_2$ values correspond to more strongly-coupled electroweak sectors.  \\
The strong sector also includes a $W^{'}_R$ boson from a $(\bf 1, \bf 3)_{0}$ composite resonance which interacts with the SM particles only after the electroweak symmetry breaking, through its electroweak mixing with the $W^{'}$ and the SM $W$. Due to its weak interaction with the SM, the $W^{'}_R$ is produced at a low rate at the LHC and is thus more difficult to discover than the $W^{'}$. We will briefly describe the $W^{'}_R$ phenomenology in App. \ref{sec:app}.\\

The low-lying fermion resonance content of the composite sector consists of the following set of vector-like quarks:

\begin{align}\label{eq:fermions} 
\begin{split}	
&	\mathcal{Q}_{2/3}=\left[\begin{array}{cc}
	T & T_{5/3} \\ 
	B & T_{2/3} \end{array}\right]=\left(\bf 2, \bf 2\right)_{2/3}  , \qquad  \tilde{T}=\left(\bf 1, \bf 1\right)_{2/3}
	\end{split}
\end{align}
\noindent
that can be arranged in a fundamental of $SO(5)$. In fact this model can describe the low-energy limit of the minimal composite Higgs model MCHM5 of Ref.~\cite{Contino:2006qr}. This is a minimal content that includes a custodial symmetry and a left-right parity to prevent large corrections to the $T$ parameter and to the $Zb\bar b$ coupling \cite{Agashe:2006at} .  \\
The composite and elementary fermions mix with each other through linear couplings \cite{Kaplan:1991dc}. In particular, the $(T,B)$ $SU(2)_L$ doublet of composite fermions mixes with a $(t^{el}_L, b^{el}_L)$ doublet in the elementary sector and the $\tilde{T}$ with an elementary $t^{el}_R$; after diagonalizing the mixings and rotating to the basis of mass eigenstates, the SM top results as a superposition of elementary and composite modes:
\begin{equation}
t_L= \cos\varphi_L t^{el}_L - \sin\varphi_L T_L
\end{equation}
and analogously for the right-handed top. $\sin\varphi_L$, that we will shortly indicate as $s_L$, represents the (left-handed) top degree of compositeness. After the EWSB, the top mass is generated as
\footnote{In order to generate also the bottom-quark mass one needs to introduce an other fundamental of SO(5), with $X=-1/3$ \cite{Bini:2011zb, Vignaroli:2012si}. The bi-doublet in this fundamental can also interact with the $W^{'}$, thus affecting its phenomenology. We will assume, however, that these resonances are heavier than the top-partner resonances in the $\bf {5}_{2/3}$. This is a reasonable assumption, considering that the bottom-partners play a less significant role in cutting-off the Higgs mass divergence. 
} 
\begin{equation}
m_t= Y_{*} s_L s_R v \ ,
\end{equation}
where $v=174$ GeV and $Y_{*}$ is the composite Higgs Yukawa coupling. In order to obtain a $\sim 174$ GeV top,  
$s_L$ must be above a minimal value of $\sim 1/Y_{*}$ -- a typical $Y_{*}$ value is $\sim 3$, giving $s^{min}_L\sim 0.33$. \\

As we will show in this study, the top partners in (\ref{eq:fermions}) play a crucial role in the $W^{'}$ phenomenology. In particular, a very promising channel for $W^{'}$ discovery, as we will show, is the $W^{'} \to T_{5/3} T_{2/3}$ channel. 
$(T_{5/3},T_{2/3})$ is the $SU(2)_L$ doublet of so-called custodians \cite{Contino:2006qr}, which are the lightest fermion resonances in the $\mathcal{Q}_{2/3}$ bidoublet. This is due to the fact that they do not directly mix with the elementary sector and thus their mass is not increased by mixing effects. Neglecting electroweak corrections, the mass of the custodians, $m_C$, is related to the mass of the $(T,B)$ doublet, $m_T$, by 
\begin{equation}
m_C= c_L \, m_T \ ,
\end{equation}
where $c_L\equiv \cos\varphi_L$. Thus, the larger is the $t_L$ degree of compositeness the larger is the mass-gap between the custodians and the other fermion resonances. \footnote{The presence of the custodians is also important to weaken the bounds from loop corrections to $T$ parameter and $Zb_L b_L$ coupling, potentially dangerous in the case of almost fully composite tops \cite{Pomarol:2008bh}.}

As discussed in the introduction, naturalness argument requires top partners below $\sim 1$ TeV.  We will thus assume in our analysis that the lightest $SU(2)_L$ doublet top-partners, the custodians, are below 1 TeV and we will fix
\begin{equation}\label{eq:mc}
m_C= 0.9 \, \text{TeV} \ .
\end{equation}
This value fulfills the indication from LHC-8 searches for $5/3$ charged vector-like quarks, which give a bound $m_C \gtrsim 0.8$ TeV on the mass of the custodians \cite{Chatrchyan:2013wfa}. The bounds on the mass of other top partners are weaker and are automatically satisfied by (\ref{eq:mc}).\\

Having introduced the model, we can now proceed to analyze the $W^{'}$ decay and production rates.\\

The $W^{'}$ boson is mainly a composite state, at least in the more strongly-coupled scenarios at large $\cot\theta_2$ (eq. (\ref{eq:rotation})). It thus couples preferentially to composite modes. In particular, the W-prime has strong interactions with the would-be-Goldstone bosons of the strong electroweak sector, $W_L Z_L$/ $W_L h$, and with the composite modes of the fermions, the top partners and, for large $s_L$ values, the SM top. On the other hand, the $W^{'}$ interacts weakly with leptons and light quarks, which are assumed to be completely elementary in the model. In particular, the interactions to light quarks, that rule the $W^{'}$ Drell-Yan production at the LHC, are inversely proportional to $\cot\theta_2$. The $W^{'}$ decay rates are the following \cite{Contino:2006nn}:

\begin{align}\label{eq:decays}
\begin{split}
 \Gamma(W^{'+}\to W^{+}_L Z_L)= \Gamma(W^{'+}\to W^{+}_L h) & =\frac{g^2_2}{192 \pi}m_{W^{'}} \cot^2\theta_2\\
\Gamma(W^{'+}\to l^{+}\nu) & =\frac{g^2_2}{48 \pi}m_{W^{'}} \tan^2\theta_2 \\
 \Gamma(W^{'+}\to \bar{q}q') & =\frac{g^2_2}{16 \pi}m_{W^{'}} \tan^2\theta_2 \\
  \Gamma(W^{'+}\to t\bar{b}) &=\frac{g^2_2}{16 \pi}m_{W^{'}} \left(s^2_L \cot\theta_2 - c^2_L \tan\theta_2 \right)^2 \\
    \Gamma(W^{'+}\to T\bar{b})= \Gamma(W^{'+}\to t \bar{B}) & =\frac{g^2_2}{16 \pi}m_{W^{'}} \frac{ s^2_L c^2_L}{\sin^2\theta_2 \cos^2\theta_2}\left(1-\frac{1}{2}\frac{m^2_T}{m^2_{W^{'}}}-\frac{1}{2}\frac{m^4_T}{m^4_{W^{'}}}  \right)\left(1-\frac{m^2_T}{m^2_{W^{'}}} \right)\\
 \Gamma(W^{'+}\to T_{5/3}\bar{T}_{2/3})& =\frac{g^2_2}{8 \pi}m_{W^{'}} \cot^2\theta_2 \left( 1+2\frac{m^2_C}{m^2_{W^{'}}}\right)\sqrt{1-4\frac{m^2_C}{m^2_{W^{'}}}}\\
  \Gamma(W^{'+}\to T\bar{B}) & =\frac{g^2_2}{16 \pi}m_{W^{'}} \Bigl\{  \Bigr . \left[ \left(c^2_L \cot\theta_2 - s^2_L \tan\theta_2 \right)^2 + \cot^2\theta_2\right]\left( 1-\frac{m^2_T}{m^2_{W^{'}}}\right)   \\
&  +6\frac{m^2_T}{m^2_{W^{'}}}\left(c^2_L \cot^2\theta_2 - s^2_L \right) \Bigl. \Bigr\} \sqrt{1-4\frac{m^2_T}{m^2_{W^{'}}}}
\end{split}
\end{align}

\begin{figure*}[tbp]
\begin{center}
\includegraphics[width=0.47\textwidth,clip,angle=0]{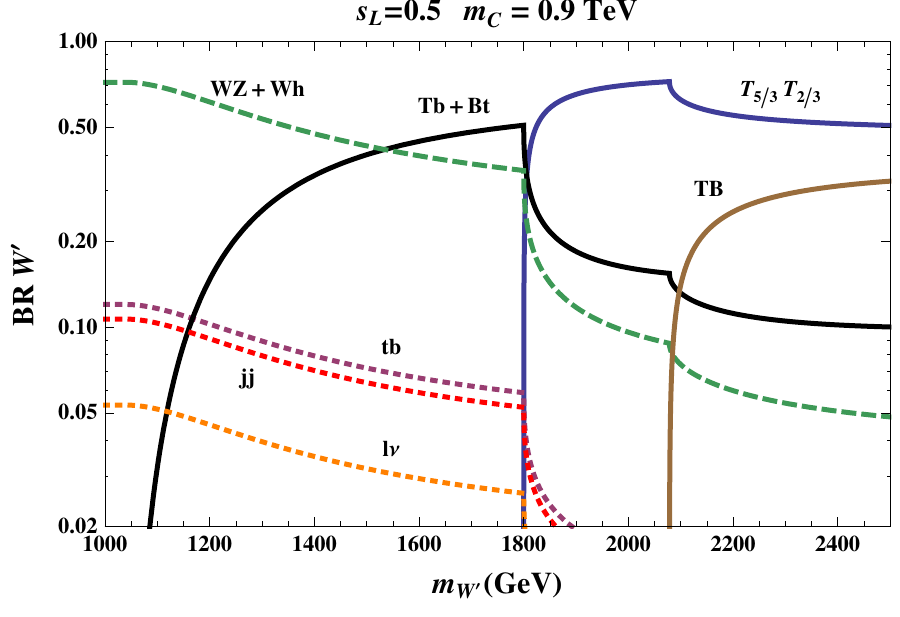}
\includegraphics[width=0.48\textwidth,clip,angle=0]{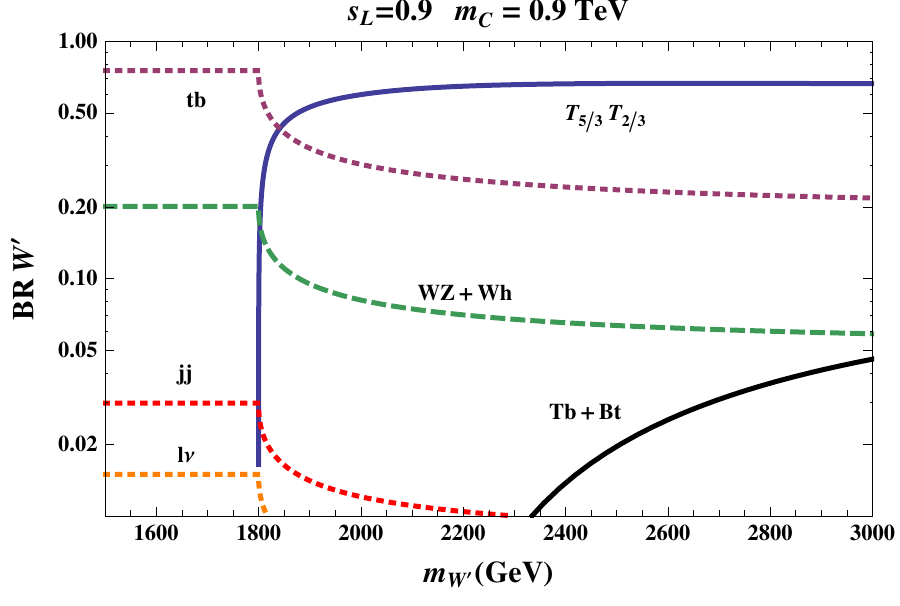}
\caption[]{
\label{fig:BR}
\small
$W^{'}$ decay branching ratios for an intermediate value $\cot\theta_2=3$ and for different top degrees of compositeness, $s_L=0.5$ (left panel) and $s_L=0.9$ (right panel). We set the custodian mass at $m_C=0.9$ TeV.
}
\end{center}
\end{figure*}

\begin{figure*}[]
\begin{center}
\includegraphics[width=0.5\textwidth,clip,angle=0]{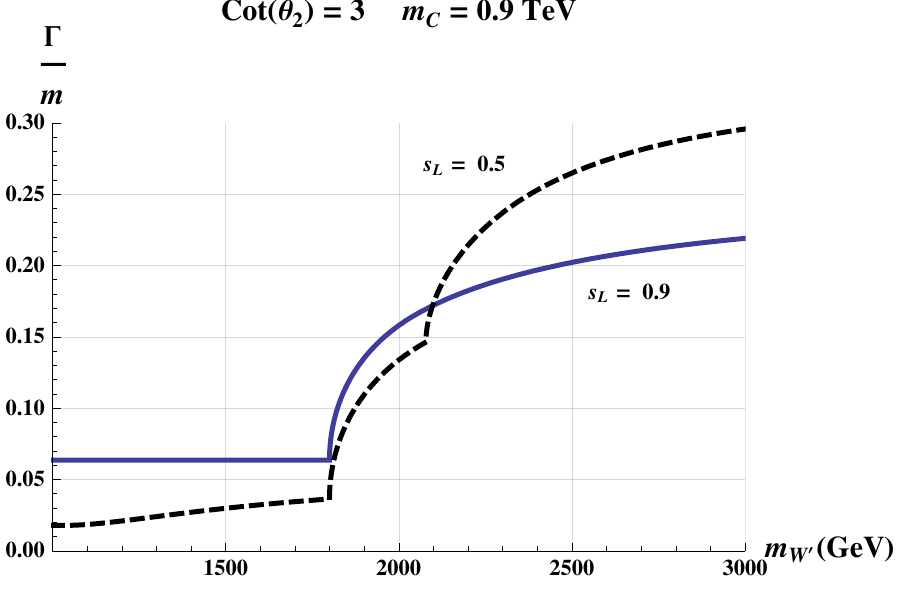}\vspace{2mm}\\
\includegraphics[width=0.38\textwidth,clip,angle=0]{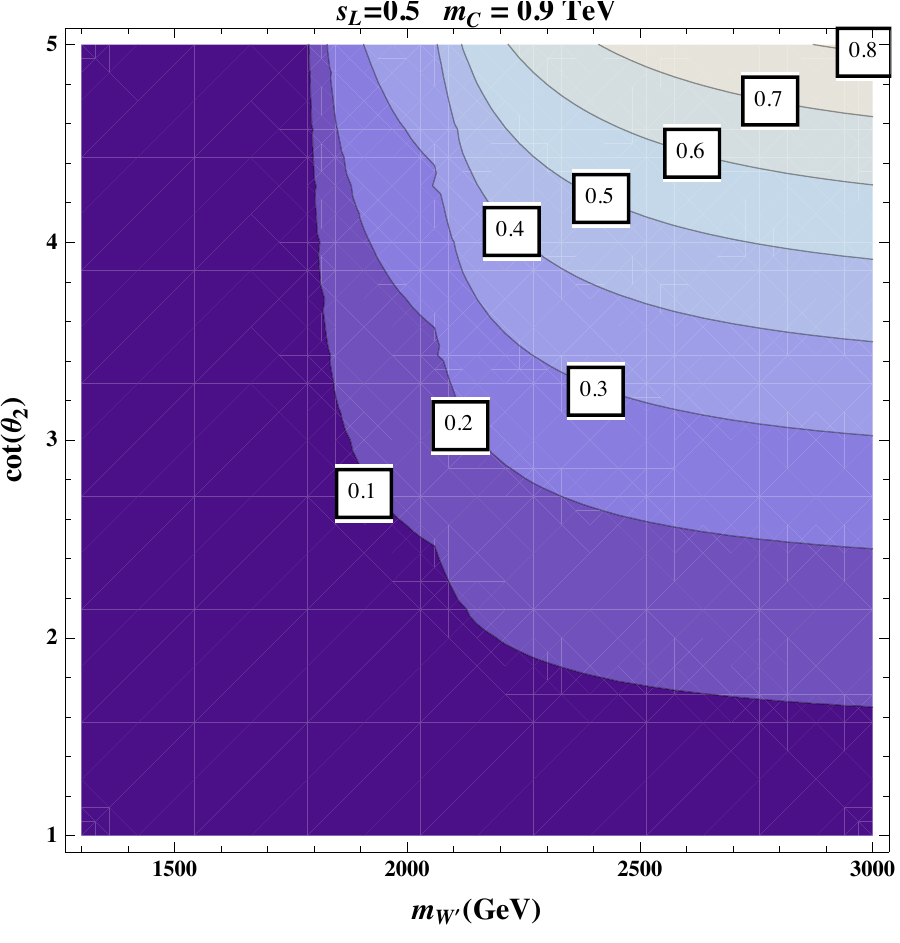} \ \ 
\includegraphics[width=0.38\textwidth, clip,angle=0]{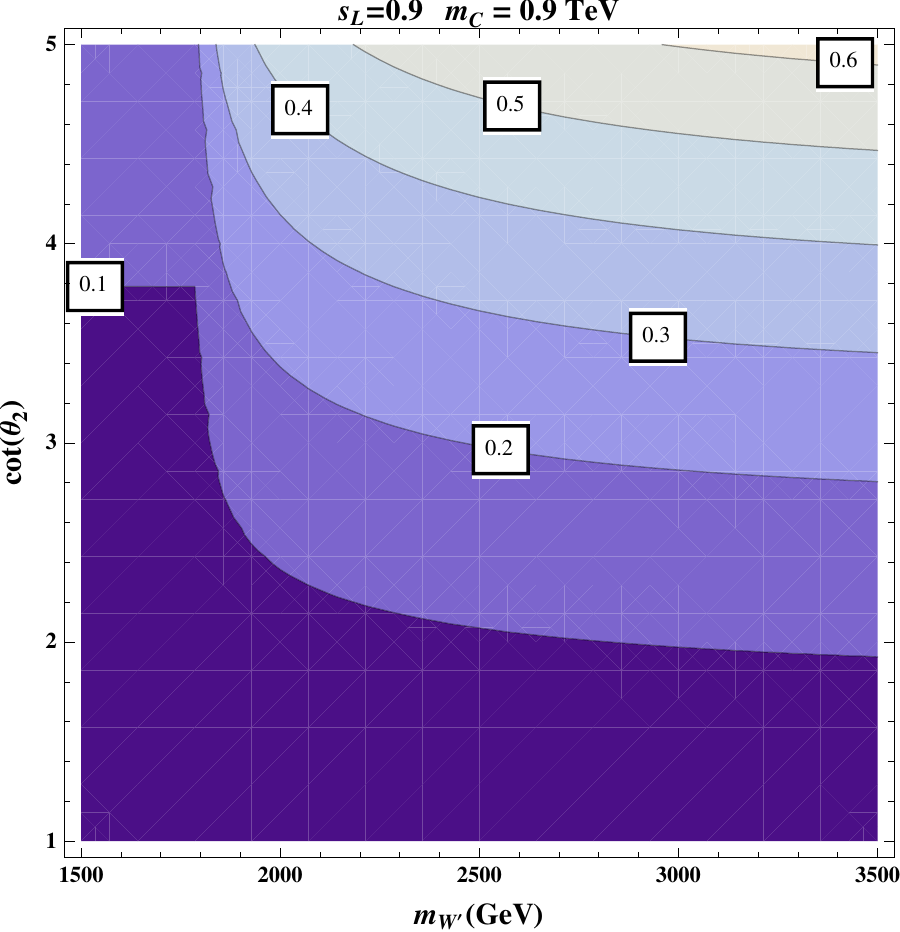}
\caption[]{
\label{fig:width}
\small
Upper panel: $W^{'}$ width-over-mass ratio for an intermediate value $\cot\theta_2=3$ and different top degrees of compositeness, $s_L=0.5, \, 0.9$. Lower panels: contour plot of the $W^{'}$ $\Gamma/m$ ratio in the $(m_{W^{'}},\cot\theta_2)$ plane for $s_L=0.5$ (left panel) and $s_L=0.9$ (right panel).
}
\end{center}
\end{figure*}

We show in Fig. \ref{fig:BR} the $W^{'}$ decay branching ratios for an intermediate value $\cot\theta_2=3$ and for different top degrees of compositeness. The plot on the right refers to a scenario of an almost fully composite top, $s_L=0.9$. We see that in this case $tb$ is the dominant decay mode in the lower mass region, $m_{W^{'}}<2 m_C$,
while above the threshold $2 m_C$ the $W^{'}$ decays almost completely into a pair of custodians. The $W^{'}\to T_{5/3}T_{2/3}$ decay is the dominant mode in the high mass region even for smaller $s_L$ values. The plot on the left of Fig. \ref{fig:BR} is obtained for $s_L=0.5$. The scenario at lower $W^{'}$ mass, instead, changes drastically with the decrease of $s_L$. We see that for a smaller $s_L=0.5$, the $tb$ branching ratio is significantly reduced; the dominant decay modes are now $WZ/Wh$ and the $W^{'}$ decays into a third generation quark plus its heavy partner, $Tb, Bt$. For $\cot\theta_2=3$ the branching ratios for the leptonic $W^{'}$ decays are quite small, below the percent level. Nevertheless the $W^{'}\to l\nu$ mode will prove to be a powerful channel to test the more weakly-coupled scenarios at lower $\cot\theta_2$ values, as we will see in the next section where we will derive the bounds set on the model parameters by LHC-8 studies.\\
The total $W^{'}$ decay width (divided by the $W^{'}$ mass) is shown in Fig. \ref{fig:width} (upper panel) for $\cot\theta_2=3$ and for $s_L=0.5,\, 0.9$. Again, we set $m_C=0.9$ TeV. The width is narrow in the lower mass region and it becomes larger, but still below $\Gamma/m\simeq 0.25-0.30$, in the heavier $W^{'}$ region above the threshold for the decay into a pair of custodians. The two lower plots in Fig. \ref{fig:width} show the $\Gamma/m$ ratio in the plane $(m_{W^{'}}, \cot\theta_2)$ for $s_L=0.5$ (left panel) and $s_L=0.9$ (right panel). We see that the width becomes quite large, with $\Gamma/m\gtrsim 0.3$, in the high mass region $m_{W^{'}}\gtrsim 2$ TeV in the more strongly-coupled regime at $\cot\theta_2\gtrsim 4$. This region of the parameter space will be thus more difficult to explore at the LHC. Nevertheless, as we will show in Sec. \ref{sec:custodian}, it can be tested through the analysis of the custodian channel. \footnote{ We will perform a signal selection for the custodian channel which do not rely on $W^{'}$ transverse mass cuts and is thus basically independent of the $W^{'}$ width.} \\

\begin{figure}[]
\begin{center}
\includegraphics[width=0.5\textwidth,clip,angle=0]{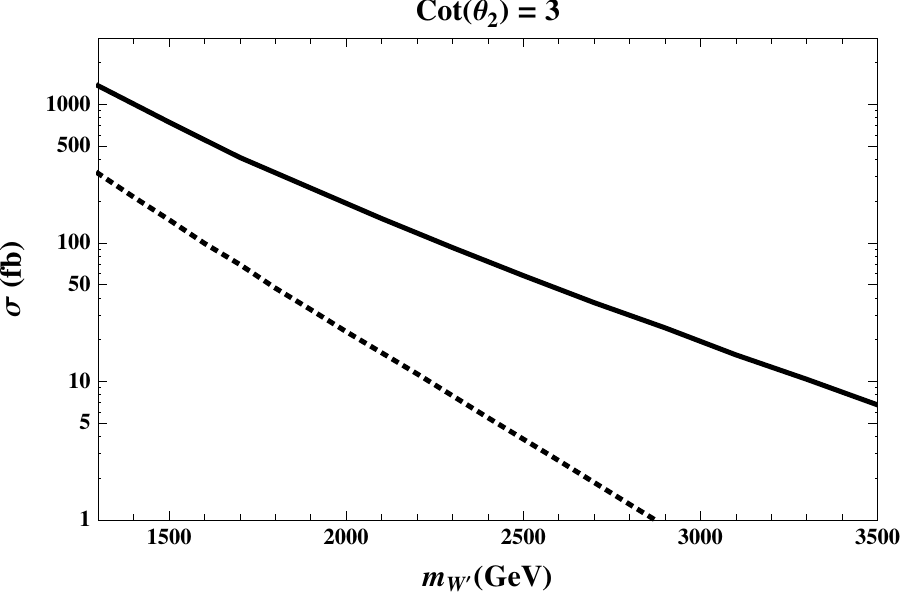}
\caption[]{
\label{fig:xsec}
\small
Total $W^{'}$ Drell-Yan cross section at the LHC with $\sqrt{s}=8$ TeV (dotted curve) and $\sqrt{s}=14$ TeV (thick curve), with $\cot\theta=3$ -- cross sections scale as $\sim 1/\cot^{2}\theta_2$.
}
\end{center}
\end{figure}

Concerning the $W^{'}$ production, Drell-Yan is the dominant $W^{'}$ production mechanism at the LHC. We show in Fig. \ref{fig:xsec} the total $W^{'}$ Drell-Yan cross section at the LHC with $\sqrt{s}=8$ TeV and $\sqrt{s}=14$ TeV and for $\cot\theta_2=3$. Due to the inverse proportionality to $\cot\theta_2$ of the $W^{'}$ couplings to light quarks, the cross section scales as $\sim 1/\cot^{2}\theta_2$. The very strongly-coupled region at $\cot\theta_2\gg 6$ will be thus difficult to explore at the 14 TeV LHC. The 8 TeV LHC can test a significant parameter region at intermediate $W^{'}$ mass, as we will also see in the next section, but has low sensitivity to the high mass region. The $m_{W^{'}}> 2 m_C$ region, on the other hand, can be extensively probed at the 14 TeV LHC.\\          
The $W^{'}$ can be also produced through (longitudinal)-weak-boson fusion. This is a relevant production mode for more strongly-coupled scenarios. However, it has been proved \cite{Agashe:2007ki, Pappadopulo:2014qza} that the weak-boson fusion 
is typically less powerful than Drell-Yan for a $W^{'}$ discovery at the LHC, at least for center-of-mass energies not exceeding 14 TeV. The $W^{'}$ production through third-generation-quark fusion is sub-leading as well, even in the scenario of fully composite tops \cite{Djouadi:2007eg}. We will thus focus our analysis on Drell-Yan production.\\

 We conclude this section by listing the most promising signatures for $W^{'}$ discovery at the LHC, as resulting from the analysis of the $W^{'}$ branching fractions. These include:
 
\begin{itemize}
\item The custodian channel $W^{'}\to T_{5/3}T_{2/3}$, in the high mass region $m_{W^{'}}>2 m_c$.
\item The heavy-light channels $W^{'}\to Tb, BT$ in the intermediate mass region and in the case of not largely composite tops.
\item $W^{'}\to W_L Z_L, W_L h$ in the intermediate mass region, especially for smaller $s_L$ values.
\item $W^{'}\to tb$ in the lower mass region but only for almost fully composite tops.
\item $W^{'}\to l\nu$ in the more weakly-coupled regime at low $\cot\theta_2$.
\end{itemize}

We will focus our analysis on the first signatures of the list, the so-far-overlooked signals of $W^{'}$ decays into vector-like quarks. We will study the custodian mode in the high $W^{'}$ mass region and, in the case of intermediate top degrees of compositeness, we will study the heavy-light channel in the lower mass region. Specifically, we will consider the $Tb$ channel. \\

Due to the different parton content of the proton,\footnote{We use the cteq6l1 pdf set \cite{Pumplin:2002vw} in all of our simulations.} the LHC Drell-Yan production cross section of a positive charged $W^{'}$ is much larger than that of a $W^{'-}$. We find $\sigma(W^{'+})\simeq 3\, \sigma(W^{'-})$ for $m_{W^{'}}\gtrsim 1.5$ TeV. For this reason, and considering that the main backgrounds to our signals will not present a so large `charge asymmetry', the analyses of Sec. \ref{sec:heavylight} and \ref{sec:custodian} will be focused on the signatures with positive charged leptons. 

\section{Current bounds from LHC-8 analyses}
\label{sec:limits}

Searches for W-prime resonances have been performed by ATLAS and CMS collaborations at the 8 TeV LHC considering, typically, a sequential-Standard-Model W-prime \cite{Altarelli:1989ff} and in the `standard' decay modes: the leptonic channel, the $WZ$ channel, the di-jet and the $tb$ modes. In this section we will extract from the results of these analyses 95$\%$ C.L. exclusion regions on the $(m_{W^{'}},\cot\theta_2)$ plane in the RS/composite Higgs model. We will derive the bounds in the narrow-width approximation, by scaling the $W^{'}$ production cross sections and decay branching fractions in accord with the values in our model. We will also include the corrections from $W^{'}-W^{'}_R$ electroweak mixing (App. \ref{sec:app}).\footnote{We will not include NLO k-factors.} As also stressed in \cite{Pappadopulo:2014qza}, the bound obtained in the narrow-width approximation are probably overestimated in the region at large $W^{'}$ width, typically at $\cot\theta_2\gtrsim 4$. In order to obtain the true limits in this large width regime one should perform dedicated analyses that take into account the correct $W^{'}$ finite width. \\ 
A derivation of the limits on a $W^{'}$ in composite Higgs models has been recently obtained in \cite{Pappadopulo:2014qza},  considering a scenario of $W^{'}$ universal couplings to fermions and with decoupled heavy fermionic resonances. Here we will consider the effect of fermionic resonances below 1 TeV, specifically we will consider $m_C=0.9$ TeV, and non-universal couplings to fermions. In particular, motivated by a scenario of partial compositeness of the SM particles, we will consider a stronger coupling of the $W^{'}$ to third-generation quarks. We will take into account different top degrees of compositeness, an intermediate degree ($s_L=0.5$) and a large degree of compositeness ($s_L=0.9$).\\

\begin{figure*}[tbp]
\begin{center}
\includegraphics[width=0.45\textwidth,clip,angle=0]{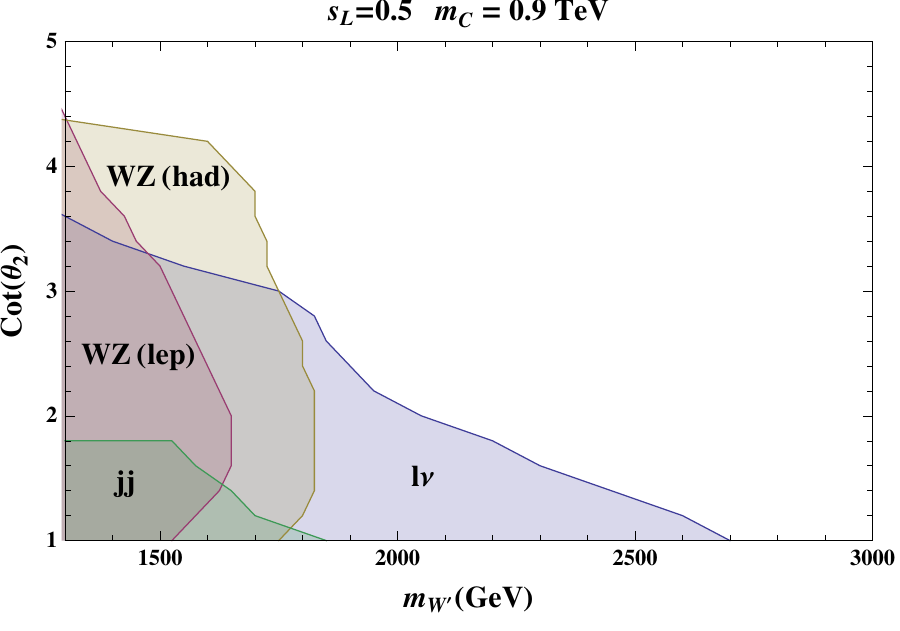}
\includegraphics[width=0.45\textwidth,clip,angle=0]{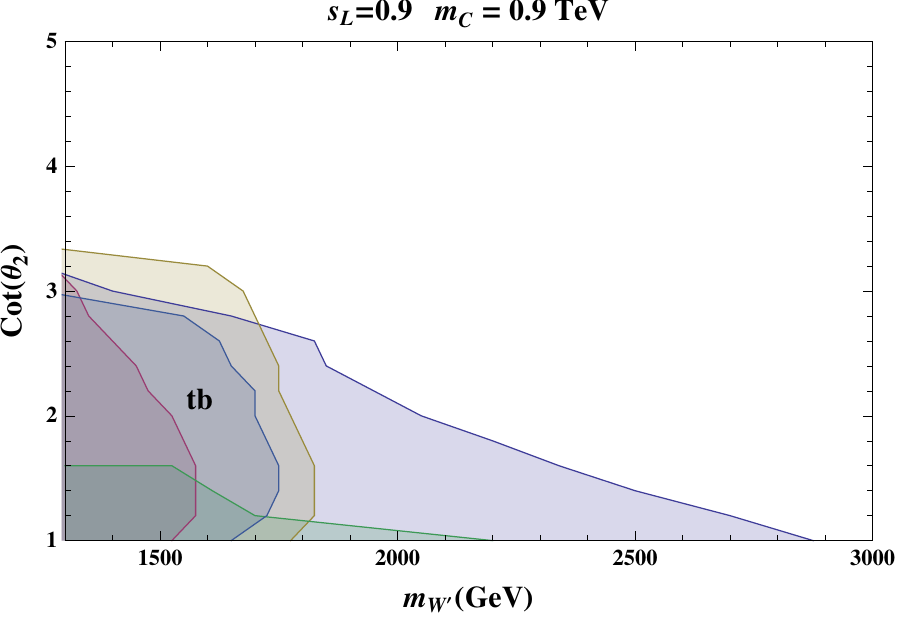}
\caption[]{
\label{fig:limits}
\small
95$\%$ C.L. excluded regions of the $(m_{W^{'}},\cot\theta_2)$ plane for $s_L=0.5$ (left plot) and $s_L=0.9$ (right plot), derived from ATLAS and CMS $W^{'}$ searches in different channels at the 8 TeV LHC.
}
\end{center}
\end{figure*}

More in details, the exclusion regions are derived from the study in \cite{CMS-PAS-EXO-12-060} of the leptonic channel $W^{'}\to l\nu$, where the lepton is an electron or a muon (we get similar results from \cite{ATLAS-CONF-2014-017}), from the analyses of the $WZ$ channel in the leptonic final state \cite{CMS-PAS-EXO-12-025} (we get similar bounds from the study in \cite{ATLAS-CONF-2014-015}) and in the hadronic final state \cite{CMS-PAS-EXO-12-024}, from the study of the di-jet channel in \cite{CMS-PAS-EXO-12-059} (similar results are obtained from \cite{ATLAS-CONF-2012-148}), where we consider a 0.6 acceptance corresponding to the case of isotropic $W^{'}$ decays, and from the analysis in \cite{CMS-PAS-b2g-12-010} (similar bounds are obtained from \cite{ATLAS-CONF-2013-050}) of the $W^{'}\to tb$ channel.  The resulting excluded regions of the $(m_{W^{'}},\cot\theta_2)$ plane are shown in Fig. \ref{fig:limits} for $s_L=0.5$ (plot on the left) and $s_L=0.9$ (plot on the right). \\

We see that the current LHC-8 analyses exclude a significant portion of the parameter space at lower $W^{'}$ masses, $m_{W^{'}}\lesssim 2 m_C$, but set much milder bounds on the high $W^{'}$ mass region, $m_{W^{'}}> 2 m_C$.
The $WZ$ channel in the hadronic final state has the largest exclusion power on the lower mass region, while the $W^{'}\to l\nu$ mode is the most powerful channel to test the more weakly-coupled regime at low $\cot\theta_2$ values. We can also observe that the bounds are less strong for higher top degree of compositeness. In this case, the $tb$ channel has the sensitivity to exclude a portion of the parameter space at lower $W^{'}$ masses.
The different exclusion potential of the diverse channels substantially confirm the expectation from the analysis of the 
$W^{'}$ decay branching ratios. We thus expect that a study of the new $W^{'}\to Tb$ mode, which is the dominant decay at lower $W^{'}$ masses and at intermediate $s_L$ values, could efficiently probe the $m_{W^{'}}\lesssim 2 m_C$ region and that the new channel of $W^{'}$ decay into custodians could extensively test the region $m_{W^{'}}> 2 m_C$. \\

In the following sections we will perform parton level analyses of the new channels where the $W^{'}$ decays into vector-like quarks, considering $m_C=0.9$ TeV. 

\section{The heavy-light decay channel }
\label{sec:heavylight}

The $W^{'}$ decay into a heavy fermion plus a SM quark, $W^{'}\to Tb, \, Bt$, is the dominant $W^{'}$ decay mode, together with the $WZ/Wh$ channel, for intermediate values of the top degree of compositeness and for relatively lighter $W^{'}$, $m_{W^{'}}\lesssim 2 m_C$. In this section we will perform a detailed analysis of the $W^{'} \to Tb$ decay channel aimed to assess the LHC discovery reach of the new signal on the plane $(m_{W^{'}}, \cot\theta_2)$ at $\sqrt{s}=14$ TeV and we will also estimate the $95\%$ C.L. exclusion region at $\sqrt{s}=8$ TeV with 20 fb$^{-1}$, that could be derived from the current LHC-8 data. \\

\begin{figure}[tbp]
\begin{center}
\includegraphics[width=0.5\textwidth,clip,angle=0]{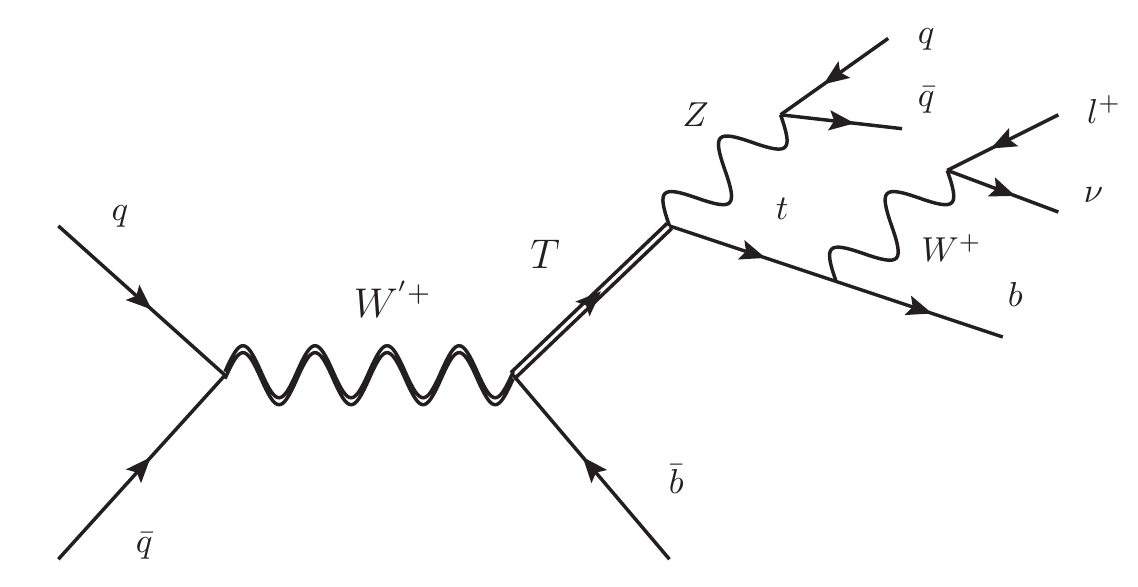}
\caption[]{
\label{fig:feyn-Tb}
\small
The $W^{' +}\to T\bar{b}$ signal. In the analysis we will also include the analogous diagram with $T\to (h\to b\bar{b}) t$ .
}
\end{center}
\end{figure}

We will start by considering a fixed $\cot\theta_2=3$ value and we set $s_L=0.5$ and $m_C=0.9$ TeV. For these values, the $T$ top partner mass and total decay width are:

\begin{equation} 
m_T=1.04 \, \text{TeV} \qquad \Gamma(T)=28\, \text{GeV} \ .
\end{equation}
\noindent
The $T$ top partner decays into $Zt$ at 49$\%$ and to $ht$, with $m_h=125$ GeV, at 51$\%$. We will include both the $(Z\to jj)t$ and the $(h\to b\bar{b})t$ decays in the analysis. \footnote{We consider a SM Higgs branching ratio for the $h\to b\bar b$ decay.} \\
Our analysis will be thus focused on the channel $pp\to W^{'+} \to (T\to ht+Zt)\, \bar{b}$,  leading to the semileptonic final state of Fig. \ref{fig:feyn-Tb}, with the lepton being an electron or a muon. \\

In our parton level analyses, the analysis of the heavy-light channel in this section and the analysis of the custodian channel in the following section, the jets will be identified with the respective quark or gluon of the tree level hard process. If the $\Delta R$ separation between two jets is below a minimum value of 0.4, we will merge the two jets in a single fat-jet. In order to mimic the effect of showering and finite detector resolution we will include a Gaussian smearing of the jet energy and momentum absolute value with $\Delta E/E = 1.0\sqrt{E/\text{GeV}}$, and of the jet momentum direction using
an angle resolution $\Delta\phi =0.05$ radians and $\Delta\eta = 0.04$. We will also include a Gaussian resolution $\sigma(\etmiss) = 0.49 \cdot \sqrt{\sum_i E^i_T/\gev}$, where $\sum_i E^i_T$ is the scalar sum of the transverse
energies of all the reconstructed objects (electrons, muons and jets), when we will compute the missing transverse energy \cite{Aad:2008zzm}.   \\

We simulate the signal events with Madgraph v.5 \cite{Alwall:2011uj}, after implementing the relevant model interactions with Feynrules \cite{Christensen:2008py}. The main backgrounds include $W^{+}W^{-}b\bar{b}$, mostly coming from $t\bar t$, that we also simulate with Madgraph, and the $Wb\bar{b}+$jets and $W+$jets backgrounds, that we simulate with ALPGEN \cite{Mangano:2002ea}. \footnote{We include all the samples with increasing multiplicity of light jets in the final state, without adopting any matching technique to remove double counting. We thus obtain a conservative estimate
of the  $Wb\bar{b}+$jets and $W+$jets backgrounds.}\\

As a first step of the analysis, we apply 
the following set of acceptance and isolation cuts on jets and leptons:
\begin{equation}
\begin{aligned}
\ptj &\geq 30\, \gev  & \quad  | \etaj | &\leq 5    & \quad 
 \Delta R_{jj} &\geq 0.4 \\[0.2cm]
\ptl &\geq 20\, \gev & \quad  | \etal | &\leq 2.5  & \quad 
 \Delta R_{jl} &\geq 0.4 \, .
\end{aligned}
\label{eq:acceptance}
\end{equation}
Where $\ptj$ ($\ptl$) and $\etaj$ ($\etal$) are respectively 
the jet (lepton) transverse momentum and pseudorapidity, and 
$\Delta R_{jj}$, $\Delta R_{jl}$  the jet-jet and jet-lepton separations. \\

In 34$\%$ of the signal events with $m_{W^{'}}=1.3$ TeV at $\sqrt{s}=14$ TeV (the percentage is almost the same for $\sqrt{s}=8$ TeV)  we reconstruct only three jets in the final state, with the percentage increasing to 45$\%$ for $m_{W^{'}}=2$ TeV. This is mainly due to the fact that the $Z$ or the $h$ bosons from the $T$ decay are boosted, being produced from the decay of heavy particles. They thus decay, for a large fraction of events, into collimated jets that are reconstructed as a single fat-jet. We do not want to loose such a large percentage of signal events. 
We thus require $n\geq 3$ jets and one positive charged lepton passing the selection (\ref{eq:acceptance}). We also require the tagging of at least two b-jet:
\begin{equation} \label{eq:evsel}
pp \to l^+ \! + n\, \text{jets} \, +  \not\!\! E_T\, , \, \,  n\geq 3 \, , \; \text{at least 2 $b$-tag}
\end{equation}
\noindent
We consider a b-tagging efficiency of 0.7 and a rejection factor of 5 for a c-jet and of 100 for a light jet \cite{b-tag}. \\

After acceptance cuts the background is by far larger than the signal, the cross sections at $\sqrt{s}=$8 TeV and 14 TeV are shown respectively in the second column of Tab. \ref{tab:8tev-1} and of Tab. \ref{tab:14tev-1}. The signal has however a very distinct topology that we can exploit to efficiently reduce the background. The main part of our selection strategy will thus 
rely on the reconstruction of the signal topology, in particular we want to reconstruct the $T$ heavy fermion, its decay products (the top and the $Z/h$ bosons), and the b-jet coming directly from the $W^{'}$. \\

\begin{figure}[]
\begin{center}
\includegraphics[width=0.4\textwidth,clip,angle=0]{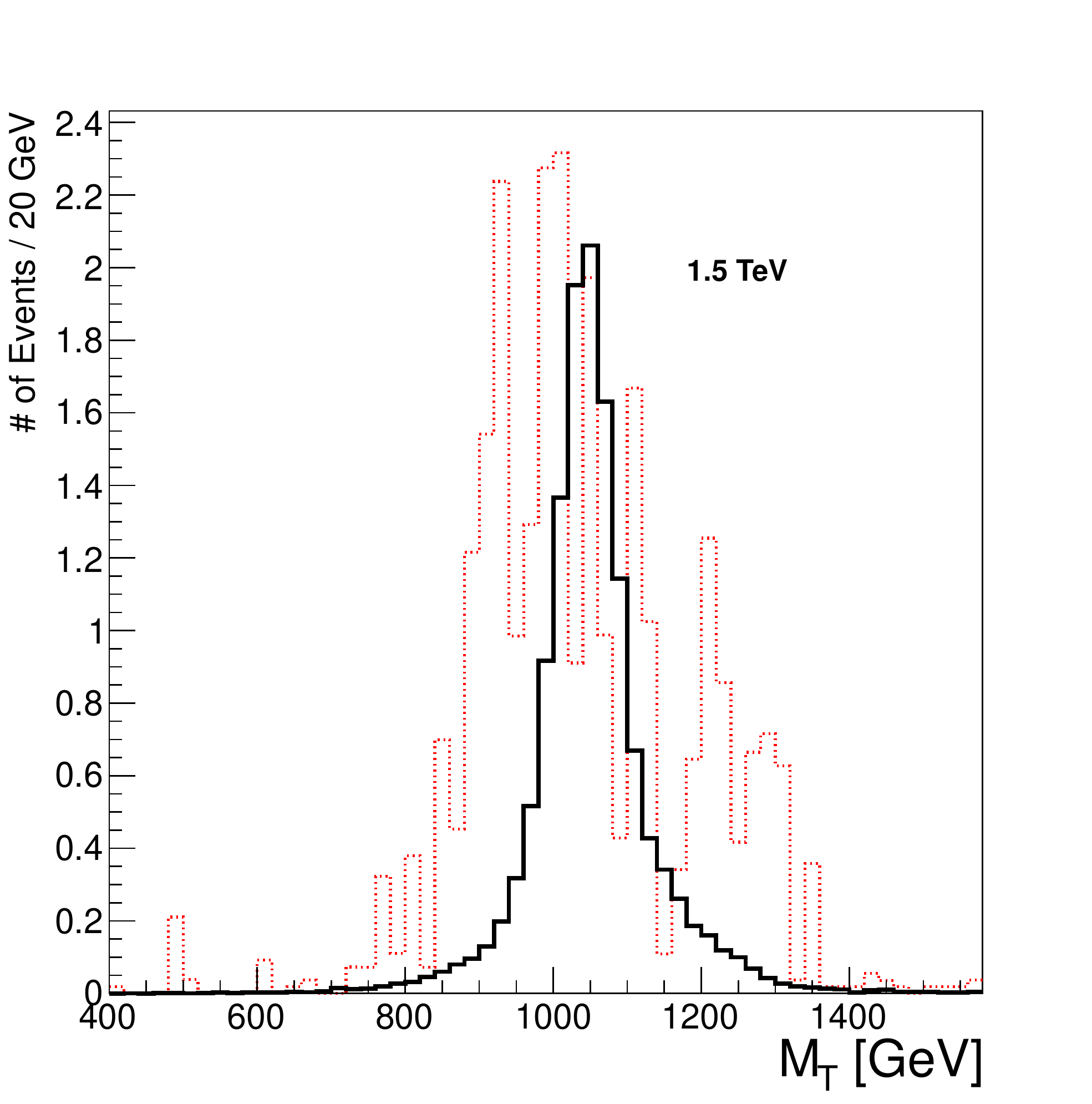}
\hspace{0.85cm}
\caption[]{
\label{fig:MT}
\small
Reconstructed $T$ invariant mass distribution for the signal $W^{'}\to Tb$ with $m_{W^{'}}=1.5$ TeV, $m_T=1.04$ TeV (solid black curve), and for the total background (red dotted curve) at the 8 TeV LHC with 20 fb$^{-1}$, after the main selection of eq. (\ref{eq:main}).
}
\end{center}
\end{figure}

The reconstruction procedure we apply goes as follows. We first reconstruct the neutrino.
The transverse momentum of the neutrino is derived by considering a $p^{TOT}_T=0$ hypothesis, as $p^{miss}_T=-\sum p_T$, where $\sum p_T$ is the sum over the $p_T$ of all the final objects (lepton and jets).
The neutrino longitudinal momentum is then obtained by requiring that the
neutrino and the lepton give an on-mass-shell $W$, $M_{l\nu} = 80.4$ GeV. 
 This condition gives two possible solutions for $p^{\nu}_z$ in $\simeq 80 \%$ of the events, both for the signal and the background \footnote{We decide to throw out the remaining $\simeq 20 \%$ of events, corresponding to the case of a quite off-shell leptonically decayed $W$, where we get imaginary solutions for $p^{\nu}_z$.   }.
Once we have reconstructed the momentum of the neutrino, we proceed to reconstruct the top coming from the $T$ decay. 
To do this, we first reconstruct the leptonically decayed $W$ (including one $W$ for each of the two $p^{\nu}_z$ solutions) and then select, among all of the possible $Wj$ combinations, the $Wj$ pair that gives the $M_{Wj}$ invariant mass closest to the top mass, $m_t=174$ GeV. This procedure also allows us to fully reconstruct the neutrino. We apply a bound on the invariant mass of the reconstructed top, $M_t \in [150,200]$ GeV, that has a $\simeq 100 \%$ efficiency on the signal and on the $WWbb$ background but that reduces the remaining backgrounds, which do not contain a top. The cross sections after the neutrino and the top reconstruction procedure at $\sqrt{s}=$8 TeV and 14 TeV are shown  in the third column of Tab. \ref{tab:8tev-1} and of Tab. \ref{tab:14tev-1} respectively. \\
The next step is the tagging of the b-jet associated with the $W^{'}$ decay. After the top reconstruction we are left with at least two possible b-jet candidates, one, barring biases in the top-tagging procedure, is the real b-jet the others are the $Z/h$ decay products. One way to distinguish the b-jet from these latter is to note that the b coming directly from the heavy $W^{'}$ is typically harder than the jets from $Z/h$. We find that this is true with the exemption of the events where the $Z/h$ decay products are merged into a single fat-jet, which is typically harder. We thus select as the b-jet the leading jet among the jets not coming from the reconstructed top if its $p_T$ is below a limit value of $(M^2_{tot}-m^2_T)/2M_{tot}$, where $M_{tot}$ is the total invariant mass; if the $p_T$ of the leading jet is above this value, which is an estimate of the b-jet $p_T$ at the Jacobian edge, the second leading jet is instead identified with the b-jet. After the b-jet tagging the $T$ top partner is nicely reconstructed, by considering as its decay products all of the final particles except the tagged b-jet. The corresponding invariant mass is shown in Fig. \ref{fig:MT} for the signal with $m_{W^{'}}=1.5$ TeV and for the total background at the 8 TeV LHC with 20 fb$^{-1}$. Signal and background distributions are obtained after the main selection we explain below.\\

The principal characteristic of our signal is the presence of heavy resonances, the $W^{'}$ and the $T$ top partner, that lead to hard final particles. Our main selection exploits this feature by applying $p_T$ cuts on the reconstructed final particles (the top, the b-jet, the $Z/h$ bosons and the heavy $T$) and a cut on $S_T$, defined as the scalar sum of the transverse momentum of all the final objects (jets, lepton and missing $p_T$). The set of cuts of the main selection consists of:

\begin{equation}
\label{eq:main}
 t \, p_T > 150\, \text{GeV} \qquad b\, p_T > 150\, \text{GeV}  \qquad Z/h \, p_T  > 150\, \text{GeV} \qquad T\, p_T  > 150\, \text{GeV} \qquad S_T > 1100\, \text{GeV}
\end{equation}

\begin{figure}[]
\begin{center}
\includegraphics[width=0.3\textwidth,clip,angle=0]{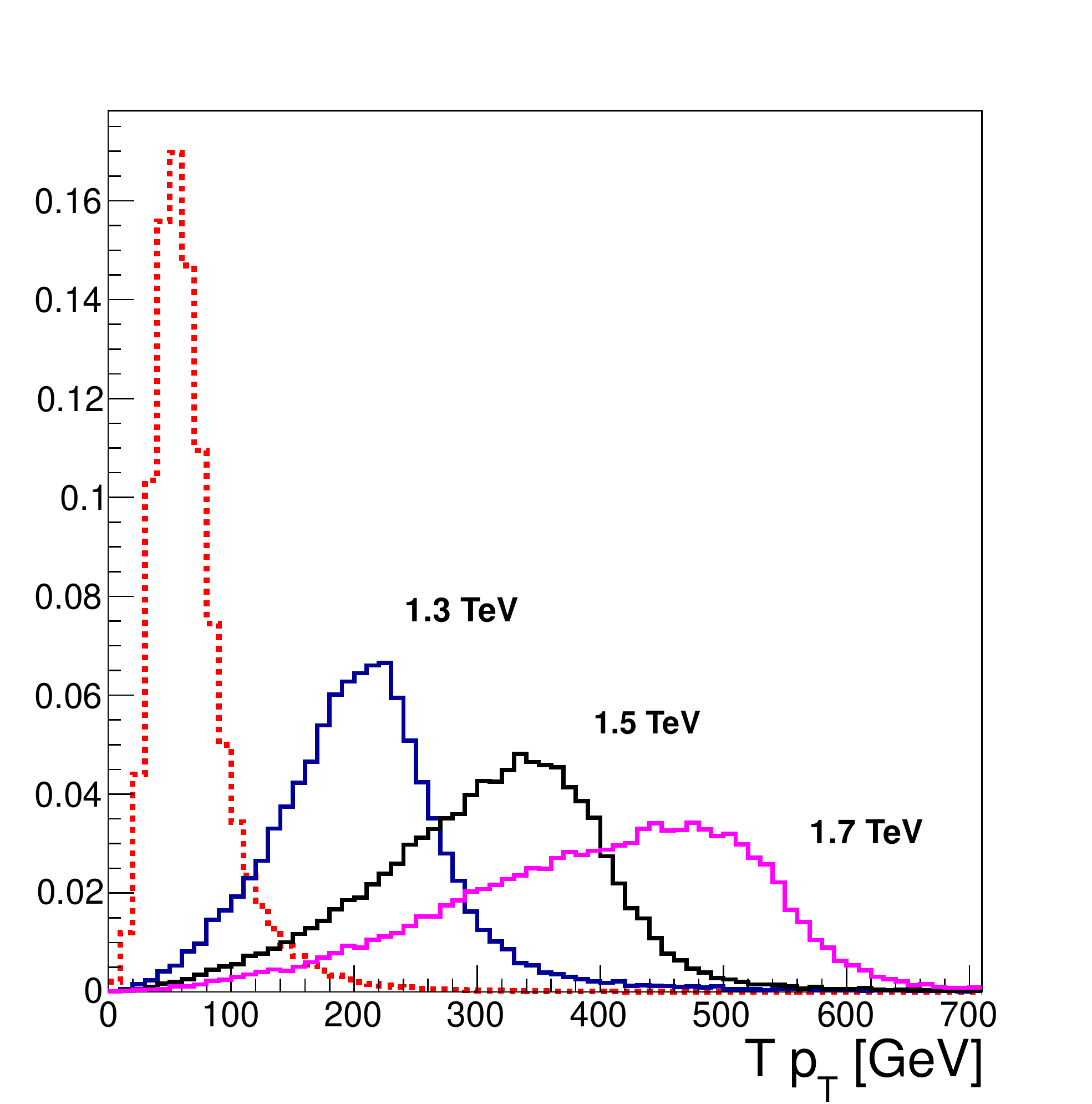}
\includegraphics[width=0.3\textwidth,clip,angle=0]{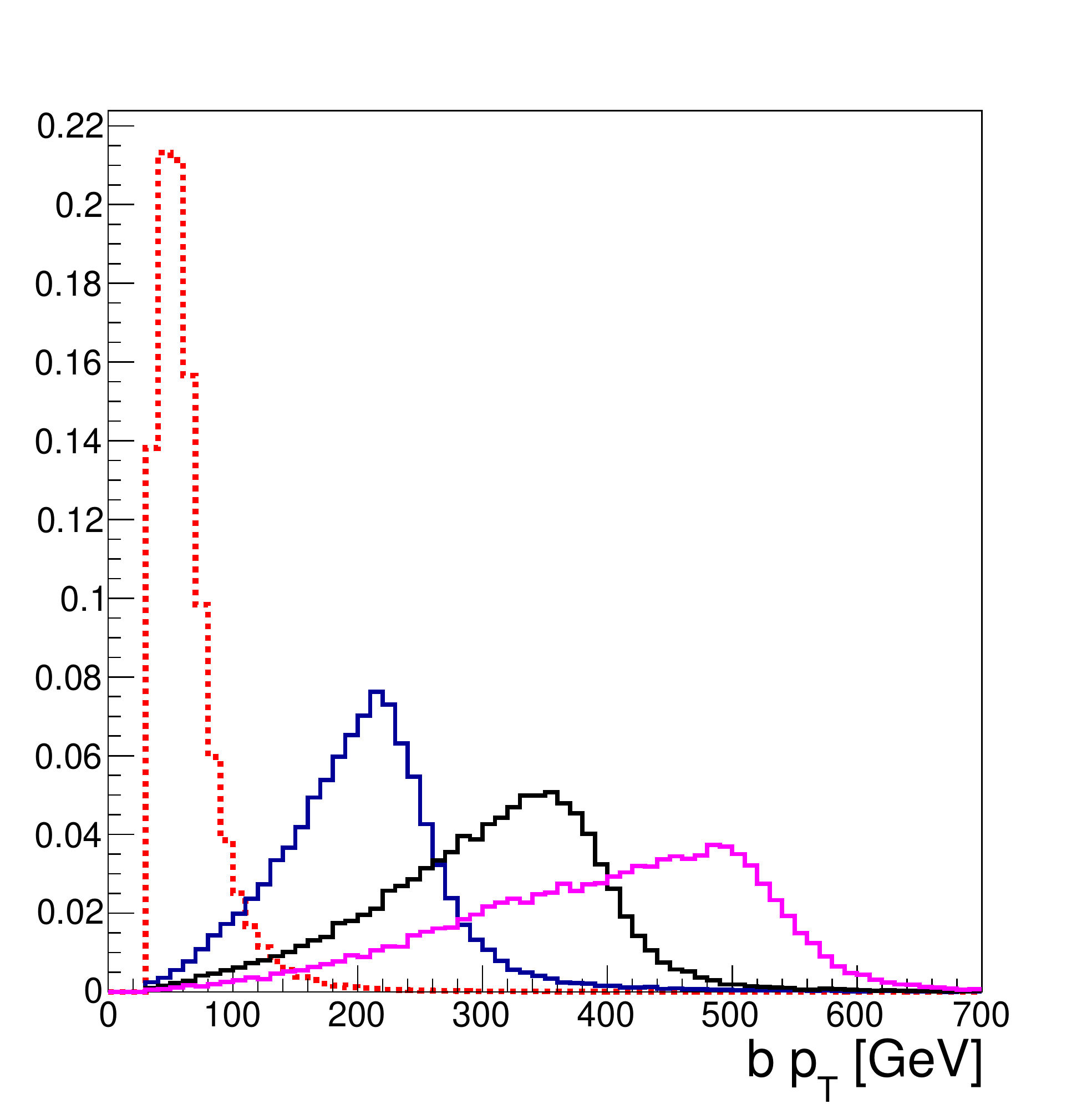}
\includegraphics[width=0.314\textwidth,clip,angle=0]{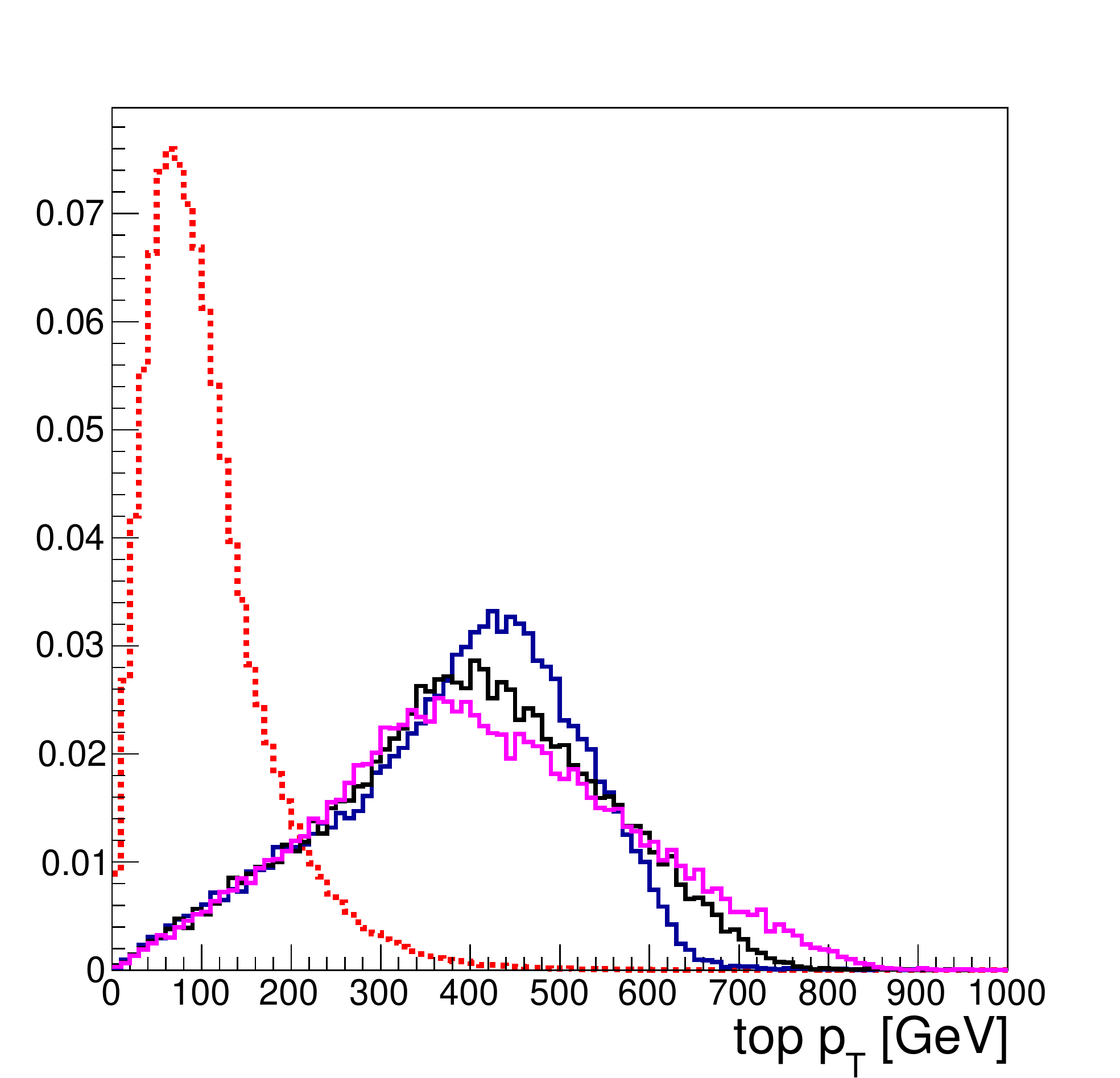}\\
\includegraphics[width=0.3\textwidth,clip,angle=0]{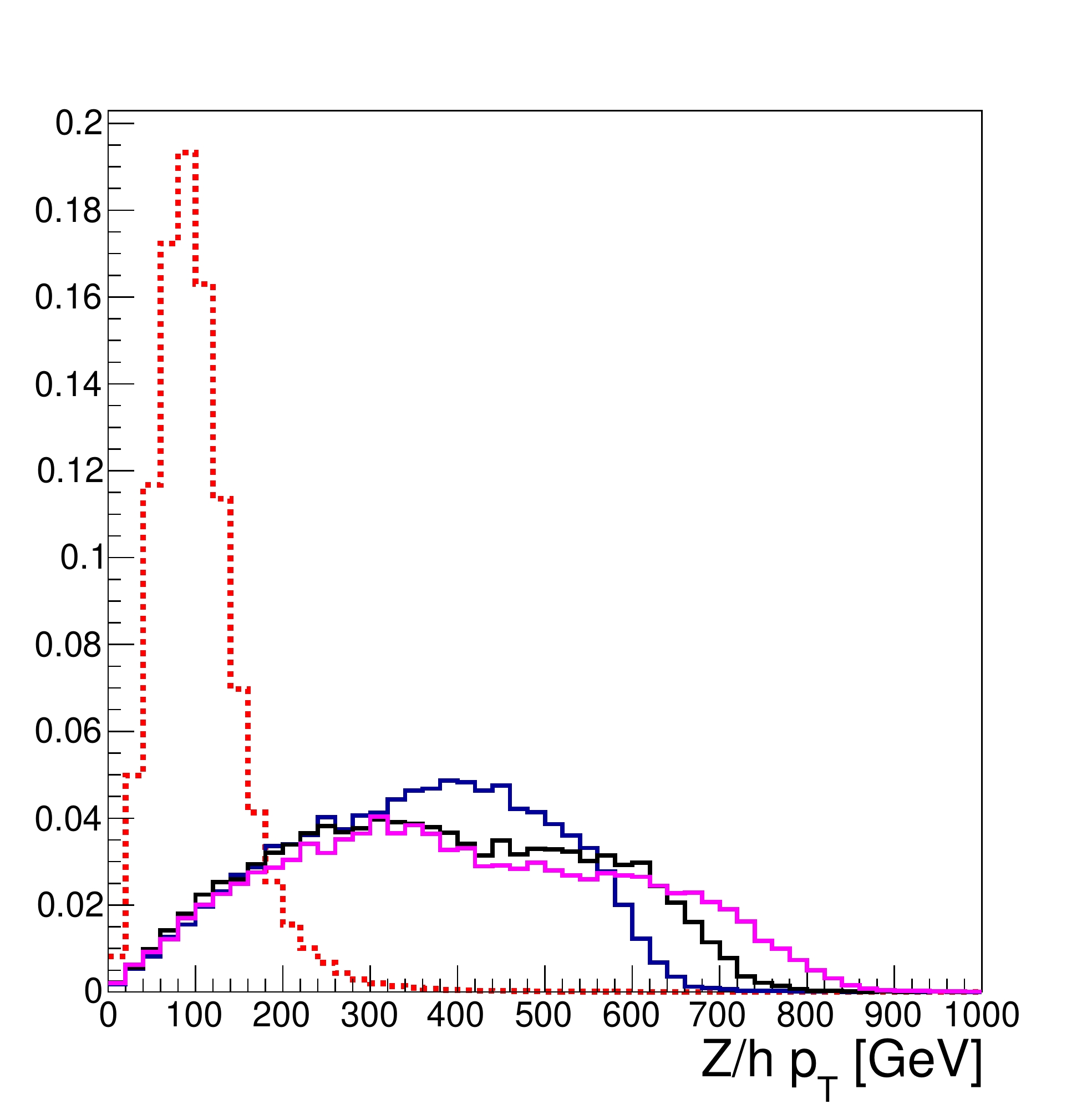}
\includegraphics[width=0.3\textwidth,clip,angle=0]{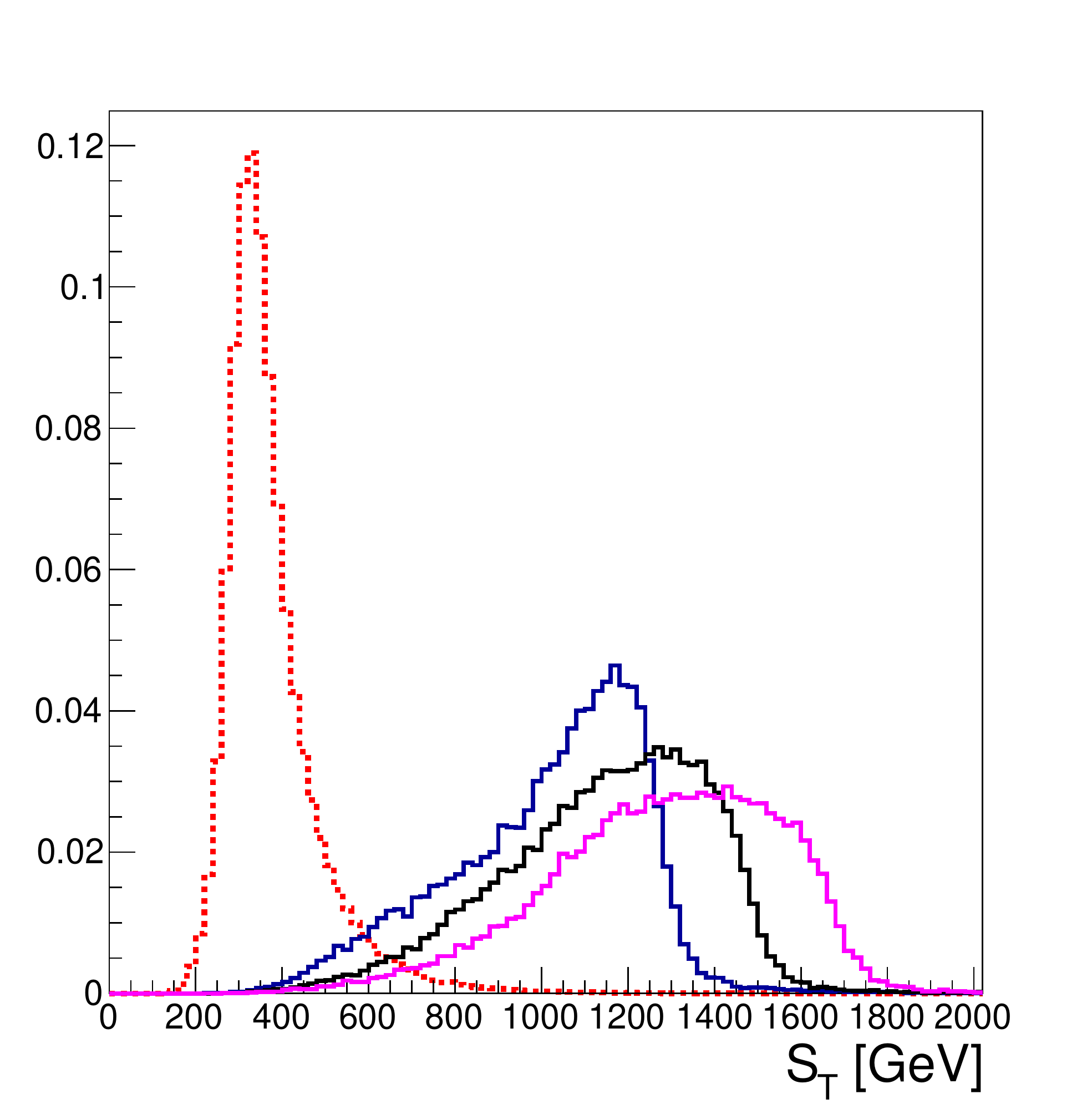}
\hspace{0.85cm}
\caption[]{
\label{fig:dist-2}
\small
$S_T$ and $p_T$ normalized distributions for the signal $W^{' +}\to T \bar{b}$ at different $W^{'}$ masses (solid curves) and the total background (red dotted curve) at the 8 TeV LHC -- very similar distributions are obtained for $\sqrt{s}=14$ TeV.
}
\end{center}
\end{figure}

\noindent
We apply the same cuts for $\sqrt{s}=8$ TeV and $\sqrt{s}=14$ TeV. \\
Fig. \ref{fig:dist-2} shows the $S_T$ and $p_T$ normalized distributions for the signal at different $W^{'}$ masses and the total background. Tab. \ref{tab:8tev-1} and Tab. \ref{tab:14tev-1} list, in the fourth column, the signal and background cross sections after the main selection for $\sqrt{s}=8$ TeV and $\sqrt{s}=14$ TeV. 
The selection is further refined by imposing a bound on the invariant mass of the reconstructed $T$, $M_{T}\in$ [0.9, 1.2] TeV; the resulting cross sections are shown in the fifth column of Tab. \ref{tab:8tev-1} and Tab. \ref{tab:14tev-1}.\\

\begin{figure}[]
\begin{center}
\includegraphics[width=0.45\textwidth,clip,angle=0]{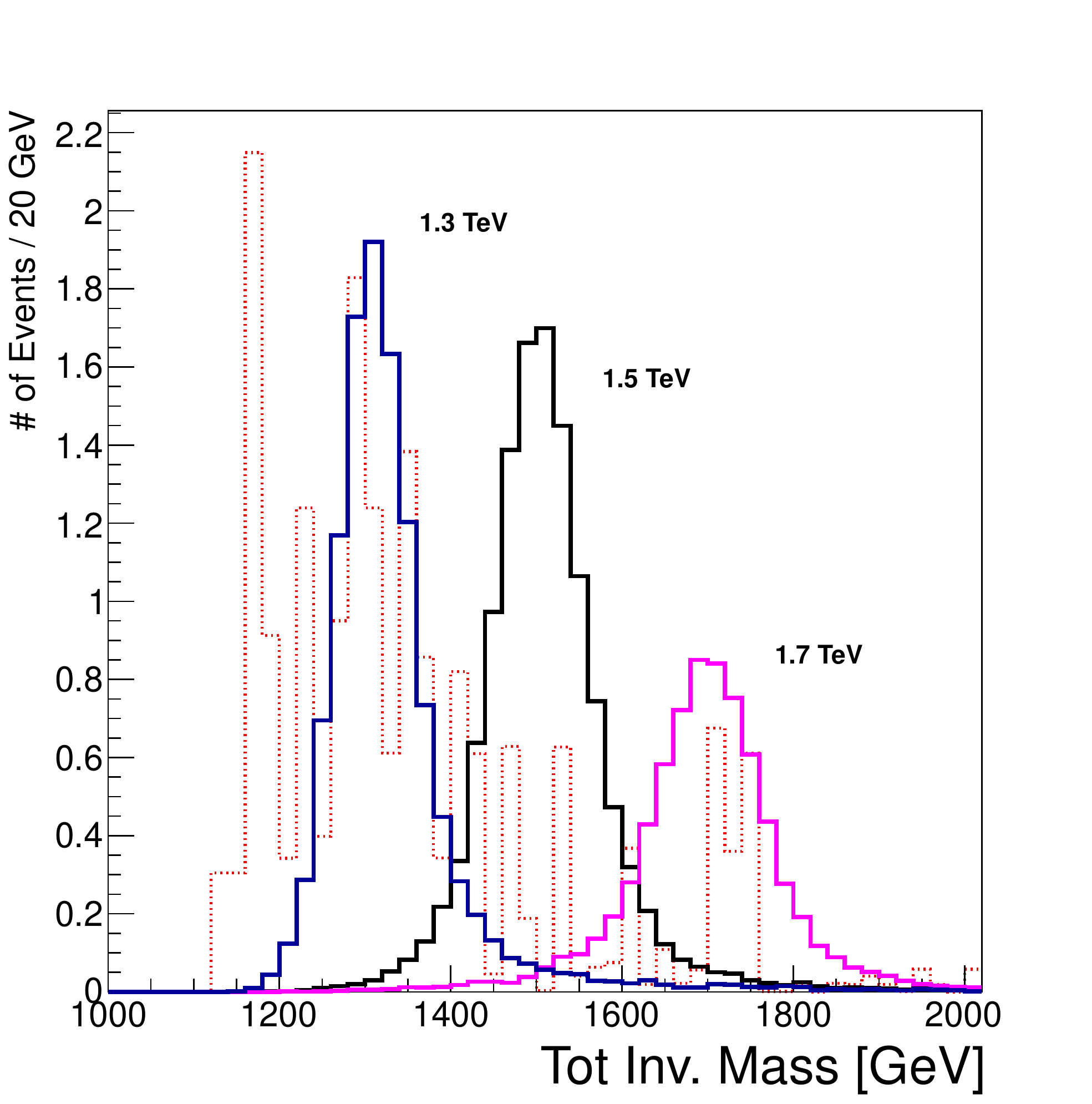}
\hspace{0.85cm}
\caption[]{
\label{fig:Mtot}
\small
Total invariant mass distribution for the signal $W^{' +}\to T \bar{b}$ at different $W^{'}$ masses (solid lines) and the total background (red dotted line) after the main selection and the cut on the $T$ invariant mass, at the LHC with $\sqrt{s}=8$ TeV and 20 fb$^{-1}$.
}
\end{center}
\end{figure}

After the main selection and the cut on the $T$ invariant mass the background is significantly reduced and we can consider the total invariant mass distribution and search for a bump in correspondence of the $W^{'}$ mass. Fig. \ref{fig:Mtot} shows the total invariant mass distribution for signals with different $W^{'}$ masses and the total background at this stage of the analysis.
We complete our selection by applying a cut on the total invariant mass. We select a region of $\pm 2\Gamma(m_{W^{'}})$ around $m_{W^{'}}$, and of $\pm 1\Gamma(m_{W^{'}})$ around $m_{W^{'}}$ in the cases with $m_{W^{'}}\geq 2$ TeV, where we fall in the large width regime. Cross section values for signal and background after the complete selection are shown in Tab. \ref{tab:8tev-fin} and Tab. \ref{tab:14tev-fin} for $\sqrt{s}=8$ TeV and $\sqrt{s}=14$ TeV. The $W+jets$ background is reduced to a negligible level and it is not shown in the tables.\\ 

\begin{figure}[]
\begin{center}
\includegraphics[width=0.65\textwidth,clip,angle=0]{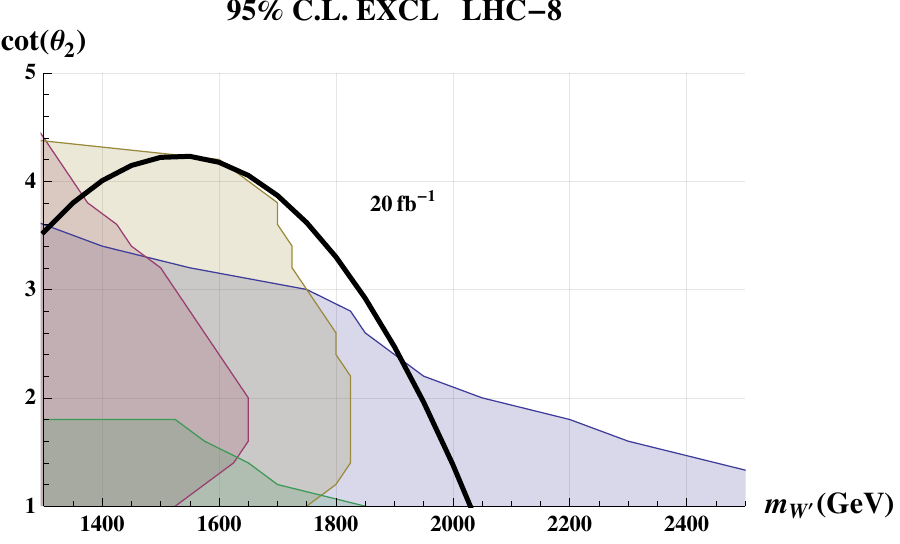}
\hspace{0.85cm}
\caption[]{
\label{fig:reach-Hl-8tev}
\small
95$\%$ C.L. exclusion reach for the channel $ pp \to \bf W^{'}\to Tb$ to semileptonic final state in the $(m_{W'}, \cot\theta_2)$ plane, with $s_L=0.5$ and $m_C=0.9$ TeV ($m_T=1.04$ TeV) at the 8 TeV LHC with 20 fb$^{-1}$. We also show the region excluded by the present LHC-8 analyses in the different `standard' $W^{'}$ decay modes (derived in Sec. \ref{sec:limits}).
}
\end{center}
\end{figure}

\begin{figure}[]
\begin{center}
\includegraphics[width=0.65\textwidth,clip,angle=0]{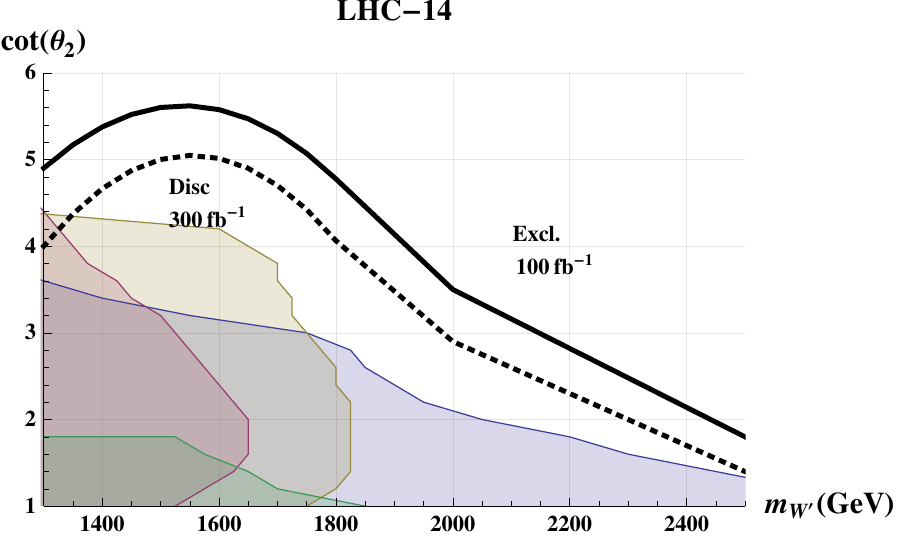}
\hspace{0.85cm}
\caption[]{
\label{fig:reach-Hl-14tev}
\small
$pp \to \mathbf{W^{'}\to Tb}$ to semileptonic final state discovery (dotted curve with 300 fb$^{-1}$) and exclusion (solid curve with 100 fb$^{-1}$) reach in the $(m_{W'}, \cot\theta_2)$ plane at the 14 TeV LHC. We set $s_L=0.5$ and $m_C=0.9$ TeV ($m_T=1.04$ TeV). We also show the region excluded by the present LHC-8 analyses in the different `standard' $W^{'}$ decay modes (derived in Sec. \ref{sec:limits}).
}
\end{center}
\end{figure}

So far we have considered a fixed value for the $W^{'}$ coupling to weak gauge bosons of eq. (\ref{eq:ct2}) $g_2\cot\theta_2=3\, g_2$. We now proceed to estimate the LHC discovery potential on the full plane $(m_{W^{'}}, \cot\theta_2)$. Starting from our results at $\cot\theta_2=3$, we consider a simple scaling of the $W^{'}$ production cross section and of the $W^{'}\to Tb$ branching ratio with $\cot\theta_2$. We also include the corrections from the $W^{'}-W^{'}_R$ electro-weak mixing, which we find to be significant in the more strongly coupled region at $\cot\theta_2 \gtrsim 4$ (more details are given in the App. \ref{sec:app}). 
The resulting LHC reach on the $(m_{W^{'}}, \cot\theta_2)$ plane is shown, together with the region excluded at 95$\%$ C.L. by the present LHC-8 analyses (derived in Sec. \ref{sec:limits}), in Fig. \ref{fig:reach-Hl-8tev} for $\sqrt{s}=8$ TeV and in Fig. \ref{fig:reach-Hl-14tev} for $\sqrt{s}=14$ TeV.\\
There is a quite complicated interplay of the $W^{'}$ production cross section, the $W^{'} \to Tb$ BR and of the background reduction efficiency trends with $\cot\theta_2$ and $m_{W^{'}}$ in explaining the estimated exclusion/discovery reach on the plane $(m_{W^{'}},\cot\theta_2)$. We find that, both for the 8 TeV and the 14 TeV LHC, the reach is well described by a parabolic curve up to $m_{W^{'}}\simeq 2$ TeV, that is up to $m_{W^{'}}$ values slightly above the $2 m_C$ threshold. The parabola at 14 TeV is somehow less peaked compared to the 8 TeV case. This is mainly caused by the effect of the EW corrections that reduce significantly the signal cross section and, as a consequence, the reach at $\cot\theta_2 \gtrsim 4$. After $\sim$ 2 TeV the reach at 14 TeV shows a tail. This is also found at the 8 TeV LHC but for $\cot\theta_2$ values below 1, which are outside the relevant parameter space of the model. The tail behavior of the reach above $\sim 2$ TeV is mainly due to the fact that the $W^{'} \to Tb$ BR is almost independent of $m_{W^{'}}$ for $m_{W^{'}}>2$ TeV and the lowering of the $W^{'}$ production cross section with $m_{W^{'}}$ is compensated by the fact that the background is reduced with a higher efficiency, by the $p_T$ cut selection and the constraint on the total invariant mass, in that region. \\
At the 8 TeV LHC with 20 fb$^{-1}$, the $Tb$ channel can test an interesting portion of the parameter space. The thick black curve in Fig. \ref{fig:reach-Hl-8tev} shows the 95$\%$ C.L. exclusion reach for the channel \footnote{We claim a 5$\sigma$ discovery if the goodness-of-fit test of
the SM-only hypothesis with Poisson distribution gives a p-value less than 2.8*10$^{-7}$ and we set a 
95$\%$ C.L. exclusion limit if the p-value of the signal plus background hypothesis is less than 0.05 }. We see that the $Tb$ mode is a very good channel to test the intermediate $W^{'}$ mass region, competitive with the $WZ$ channel in the hadronic final state, and that it could even slightly extend the current bounds from LHC-8 analyses. We point out that these latter limits have been derived in a conservative way, by considering the narrow width approximation, and that reasonably they overestimate the true bounds in the large $\cot\theta_2 \gtrsim 4$ region.  The 5$\sigma$ discovery reach with 20 fb$^{-1}$ at 8 TeV is found to be completely within the region already excluded by other searches.\\
The discovery and exclusion reach of the $Tb$ channel is extended at the 14 TeV LHC. With 100 fb$^{-1}$, thick curve in Fig. \ref{fig:reach-Hl-14tev}, a study of the channel can almost completely exclude the hypothesis of a $W^{'}$ in the intermediate mass range in the case of a not largely composite top.  While a discovery in a large portion of the parameter space not excluded by the LHC-8 studies could occur with about 300 fb$^{-1}$, dotted curve in Fig. \ref{fig:reach-Hl-14tev}.

\begin{table}
\begin{center}
{\small
\begin{tabular}{|c|c|c|c|c|}
\hline 
 & & & &  \\
 & & & &  \\[-0.6cm]
 \textsf{LHC-8} & acceptance & $\nu$+$t$ reco & MAIN sel. & MAIN sel. +$M_{T}$ cut \\  [0.2cm]
 \hline
  & & & &  \\[-0.6cm]
  & & & & \\
 $m_{W^{'}}=1.3$ TeV  & 2.3 & 1.7 & 0.59 &  0.56        \\[0.25cm]
 $m_{W^{'}}=1.5$ TeV   & 1.6 & 1.2 & 0.66 &   0.61        \\[0.25cm]
  $m_{W^{'}}=1.7$ TeV   & 0.88 & 0.64 & 0.44 &  0.36      \\[0.25cm]
   $m_{W^{'}}=2.0$ TeV   & 0.08 & 0.05 & 0.04  &  0.03      \\[0.25cm]
\hline
 & & & & \\
 & & & &  \\[-0.6cm]
  $WWbb$            & 5000 & 3800 & 1.2 & 0.84 \\[0.17cm]
  $Wbb+jets$        & 500 & 230 & 0.22 &  0.10    \\[0.17cm]
  $W+jets$            & 130 & 61 & 0.05 &  0.01 \\[0.2cm]
  Total                  & & & &  \\
  background      &   5600 &  4100 &  1.5  & 0.95    \\[0.15cm]
\hline
\end{tabular}
}
\caption{
\label{tab:8tev-1}
\small 
Cross sections, in fb, at $\sqrt{s}=8\,$TeV for the signal $W^{' +}\to T \bar{b}$ and the main backgrounds after imposing the acceptance cuts of 
eq.(\ref{eq:evsel}), including the b-tagging efficiency and rejection rates (second column); after the reconstruction procedure, including the neutrino and the top reconstruction efficiencies (third column), after the main selection of eq. (\ref{eq:main}) (fourth column) and after the further restriction on the $T$ invariant mass, $M_{T}\in$ [0.9, 1.2] TeV (fifth column). We set $\cot\theta_2=3$.
}
\end{center}
\end{table}

\begin{table}
\begin{center}
{\small
\begin{tabular}{|c|c|c|c|c|}
\hline 
 & & & &  \\
 & & & &  \\[-0.6cm]
 \textsf{LHC-14} & acceptance & $\nu$+$t$ reco & MAIN sel. & MAIN sel. +$M_{T}$ cut \\  [0.2cm]
 \hline
  & & & &  \\[-0.6cm]
  & & & & \\
 $m_{W^{'}}=1.3$ TeV  & 9.1 & 6.7 & 2.4 & 2.3          \\[0.25cm]
 $m_{W^{'}}=1.5$ TeV   & 7.5 & 5.6 & 3.2 & 2.9          \\[0.25cm]
  $m_{W^{'}}=1.7$ TeV   & 5.0 & 3.7  & 2.6 & 2.1        \\[0.25cm]
   $m_{W^{'}}=2.0$ TeV   & 0.70 & 0.51 & 0.39 & 0.30        \\[0.25cm]
    $m_{W^{'}}=2.5$ TeV   & 0.11 & 0.07 & 0.05 & 0.04        \\[0.25cm]
\hline
 & & & & \\
 & & & &  \\[-0.6cm]
  $WWbb$           & 19000 & 14000 & 11 & 7.6   \\[0.17cm]
  $Wbb+jets$        & 1600 & 920 & 2.6  &  1.0    \\[0.17cm]
  $W+jets$            & 560 & 260 & 0.66 &  0.06 \\[0.25cm]
  Total                  & & & &  \\
  background      & 21000  &  15000 &  14 &  8.7   \\[0.15cm]
\hline
\end{tabular}
}
\caption{
\label{tab:14tev-1}
\small 
Cross sections, in fb, at $\sqrt{s}=14\,$TeV for the signal $W^{' +}\to T \bar{b}$ and the main backgrounds after imposing the acceptance cuts of 
eq.(\ref{eq:evsel}), including the b-tagging efficiency and rejection rates (second column); after the reconstruction procedure, including the neutrino and the top reconstruction efficiencies (third column), after the main selection of eq. (\ref{eq:main}) (fourth column) and after the further restriction on the $T$ invariant mass, $M_{T}\in$ [0.9, 1.2] TeV (fifth column). We set $\cot\theta_2=3$.
}
\end{center}
\end{table}

\begin{table}
\begin{center}
{\small
\begin{tabular}{|c|c|c|c|c|}
\hline 
 & & & &  \\
 & & & &  \\[-0.6cm]
 \textsf{LHC-8} & $m_{W^{'}}=1.3$ TeV & $m_{W^{'}}=1.5$ TeV & $m_{W^{'}}=1.7$ TeV & $m_{W^{'}}=2.0$ TeV \\  [0.2cm]
 \hline
  & & & &  \\[-0.6cm]
  & & & & \\
  $W^{' +}\to T \bar{b}$  & 0.42 & 0.50 & 0.31 &  0.024        \\[0.25cm]
 & & & & \\
 & & & &  \\[-0.6cm]
  $WWbb$            & 0.31 & 0.11 & 0.09 & 0.046 \\[0.17cm]
  $Wbb+jets$        & 0.02 & 0.02 & 0.02 &  0.014    \\[0.17cm]
  Total                  & & & &  \\
  background      &   0.33 &  0.13 &  0.11  & 0.060    \\[0.15cm]
\hline
\end{tabular}
}
\caption{
\label{tab:8tev-fin}
\small 
Cross sections, in fb, at $\sqrt{s}=14\,$TeV for the signal $W^{' +}\to T \bar{b}$ (with $\cot\theta_2=3$) and the main backgrounds after the complete selection. 
}
\end{center}
\end{table}

\begin{table}
\begin{center}
{\small
\begin{tabular}{|c|c|c|c|c|c|}
\hline 
 & & & & & \\
 & & & & & \\[-0.6cm]
 \textsf{LHC-14} & $m_{W^{'}}=1.3$ TeV & $m_{W^{'}}=1.5$ TeV & $m_{W^{'}}=1.7$ TeV & $m_{W^{'}}=2.0$ TeV & $m_{W^{'}}=2.5$ TeV \\  [0.2cm]
 \hline
  & & & & & \\[-0.6cm]
  & & & & & \\
  $W^{' +}\to T \bar{b}$  & 1.7 & 2.4 & 1.8 &  0.24 & 0.04        \\[0.25cm]
 & & & & & \\
 & & & &  & \\[-0.6cm]
  $WWbb$            & 2.4 & 1.2 & 1.0 & 0.63 & 0.49 \\[0.17cm]
  $Wbb+jets$        & 0.1 & 0.2 & 0.2 &  0.21 & 0.20    \\[0.17cm]
  Total                  & & & & & \\
  background      &   2.5 &  1.4 &  1.2  & 0.84 & 0.69    \\[0.15cm]
\hline
\end{tabular}
}
\caption{
\label{tab:14tev-fin}
\small 
Cross sections, in fb, at $\sqrt{s}=14\,$TeV for the signal $W^{' +}\to T \bar{b}$ (with $\cot\theta_2=3$) and the main backgrounds after the complete selection. 
}
\end{center}
\end{table}

\section{The custodian channel in the same-sign dilepton final state }
\label{sec:custodian}

\begin{figure}[t]
\begin{center}
\includegraphics[width=0.5\textwidth,clip,angle=0]{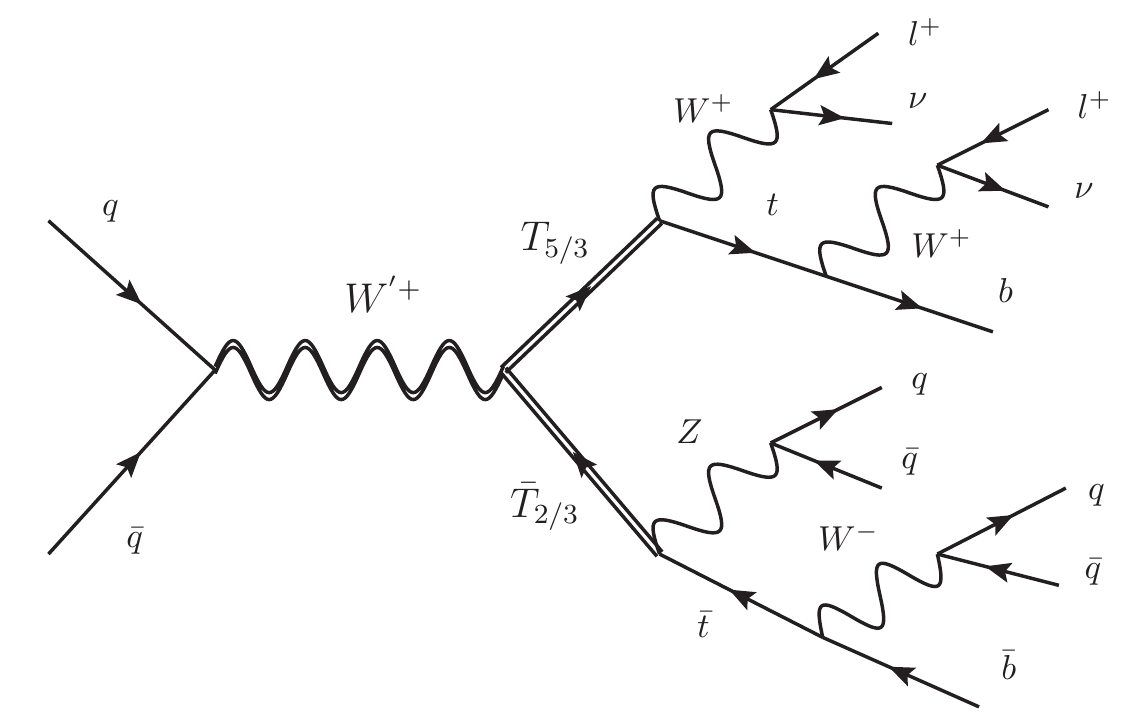}
\caption[]{
\label{fig:feyn-cust}
\small
The $W^{' +}\to T_{5/3}\bar{T}_{2/3}$ to same-sign dilepton signal. We will also consider the analogous channel with $\bar{T}_{2/3}\to (h\to b\bar{b})\bar{t}$ .
}
\end{center}
\end{figure}

The region at high $W^{'}$ mass, $m_{W^{'}}\gtrsim 2 m_C$, is favored by both naturalness argument and electro-weak-precision tests \cite{Ciuchini:2013pca, Grojean:2013qca}. Due to a lower $W^{'}$ production cross section, this region is basically not accessible at the 8 TeV LHC (except for a thin region at lower $\cot\theta_2$ values), but it could be largely explored at the 14 TeV LHC through the study of the $W^{'}$ decay into a doublet of custodian heavy fermions.
In this section we will perform a detailed analysis of the $pp\to W^{'+}\to T_{5/3}\bar{T}_{2/3}$ channel in the same-sign dilepton final state (Fig. \ref{fig:feyn-cust}).  The very clean same-sign dilepton signature is indeed one of the best channels to study,  
considering that, in spite of a low rate, 
we can have enough statistics with the expected large integrated luminosity at the 14 TeV LHC.  \\ 

 We consider a large top degree of compositeness, $s_L=0.9$, and $m_C=0.9$ TeV. \footnote{We set $m_{T_{5/3}}=m_{T_{2/3}}=m_C$. We find indeed that the correction to $m_{T_{2/3}}$ induced by the $T_{2/3}$ electro-weak mixing with other $2/3$ charged fermions is very small for large $s_L$ values. We find a order 1$\%$ correction for $s_L=0.9$.} For those values, the custodian total decay widths are:

\begin{equation} 
\Gamma(T_{5/3})=9.7\, \text{GeV}  \qquad \Gamma(T_{2/3})=11\, \text{GeV} \ .
\end{equation}
\noindent
$T_{5/3}$ decays completely into $Wt$; $T_{2/3}$ decays 48$\%$ into $Zt$ and 52$\%$ to $ht$, with $m_h=125$ GeV. We will include both the $(Z\to jj)t$ and the $(h\to b\bar{b})t$ decays in the analysis. \\

We simulate signals and backgrounds with Madgraph. The main backgrounds include $W^{+} t\bar t+$ jets, $W^{+}W^{+}+$ jets and $W^{+}W^{+}W^{-}+$ jets. \footnote{We include all the samples with increasing multiplicity of light jets in the final state, up to the highest jet multiplicity that is possible to simulate with Madgraph. As in the previous analysis, we do not apply matching procedures to remove double counting and we thus expect to obtain a conservative estimate of the background. 
} We have checked that other backgrounds, as $WWt \bar t$, $WWZ$, $WZt \bar t$ and $Wht \bar t$, are very small after acceptance cuts and are reduced to a negligible level after the main selection that we will apply. \\

As a first step of the analysis we impose the set of acceptance and isolation cuts on jets and leptons of eq. (\ref{eq:acceptance}). 
We find that, for the majority (around 40$\%$) of the signal events we reconstruct five jets in the final state. We find also a large percentage of events, ranging from 23$\%$ at $m_{W^{'}}=1.9$ TeV to 34$\%$ at $m_{W^{'}}=3.0$ TeV, with only four final reconstructed jets. This is mainly due to the merging in a single fat-jet of the particles produced by the $Z/h$ and/or the top decay; where $Z/h$ and the hadronically decaying top are the $T_{2/3}$ decay products, which are typically highly boosted. Considering this large fraction of signal events with four final jets,      
we will require in our analysis $n\geq 4$ jets and two positive charged lepton passing the selection (\ref{eq:acceptance}):
\begin{align} \label{eq:evsel-cust}
\begin{split}
& pp \to l^+ l^+ \! + n\, \text{jets} \, +  \not\!\! E_T\, , \qquad n\geq 4 \qquad  l^{+} l^+=e^{+}e^+,\mu^+\mu^{+}, e^+\mu^+\ .
\end{split}
\end{align}

\begin{figure}[]
\begin{center}
\includegraphics[width=0.3\textwidth,clip,angle=0]{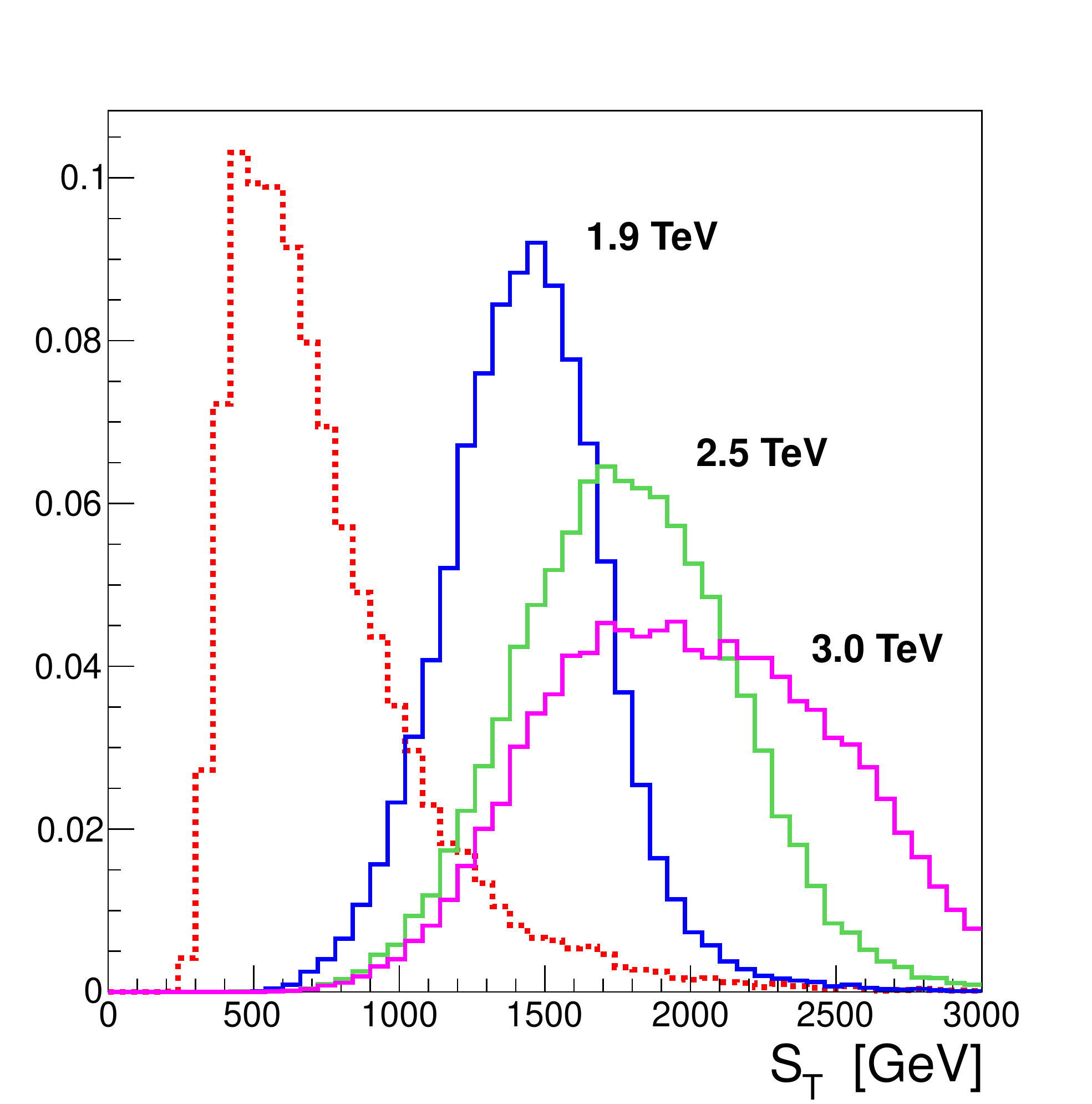}
\includegraphics[width=0.3\textwidth,clip,angle=0]{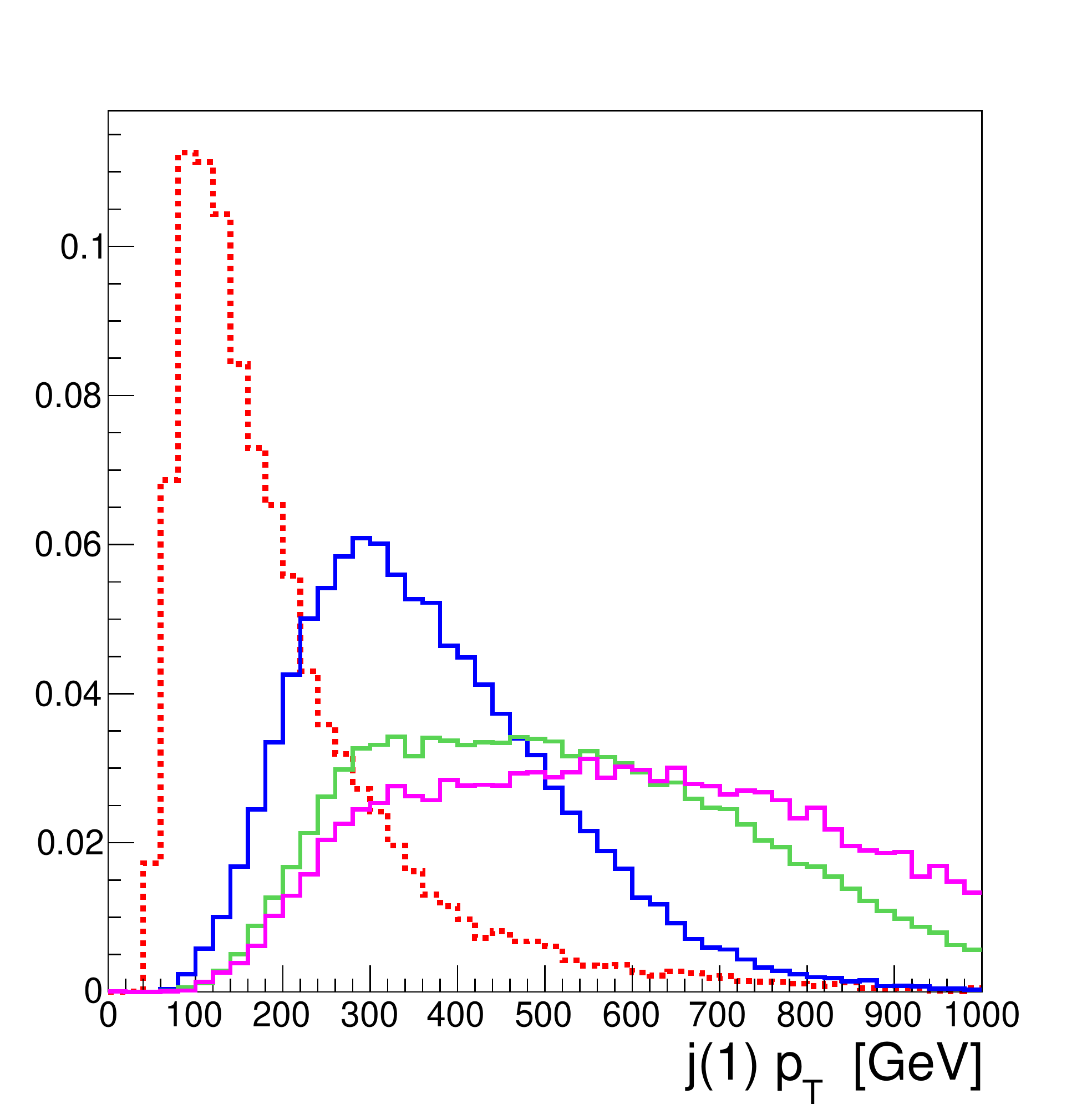}
\includegraphics[width=0.315\textwidth,clip,angle=0]{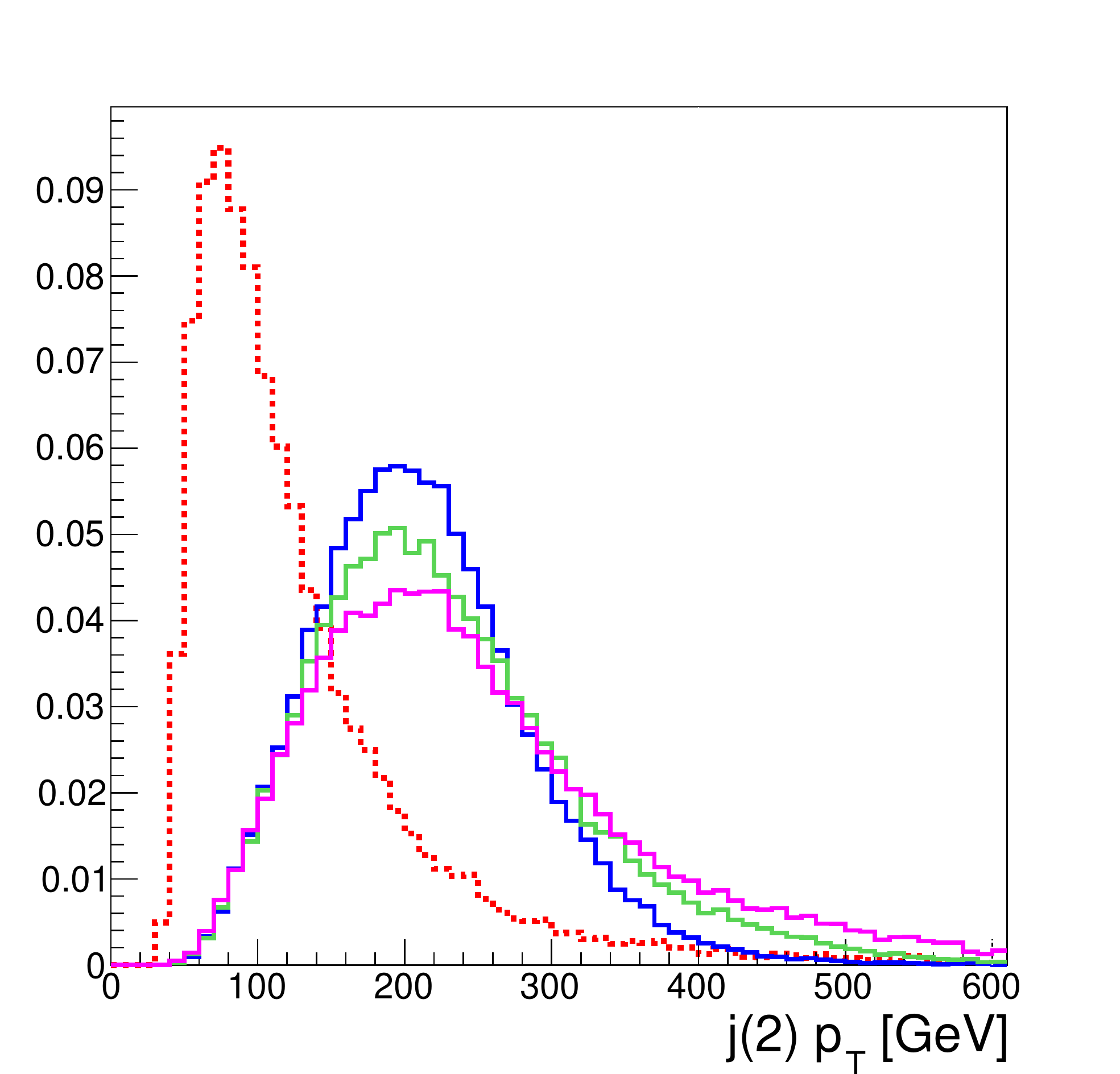}\\
\includegraphics[width=0.3\textwidth, height=0.29\textwidth, clip,angle=0]{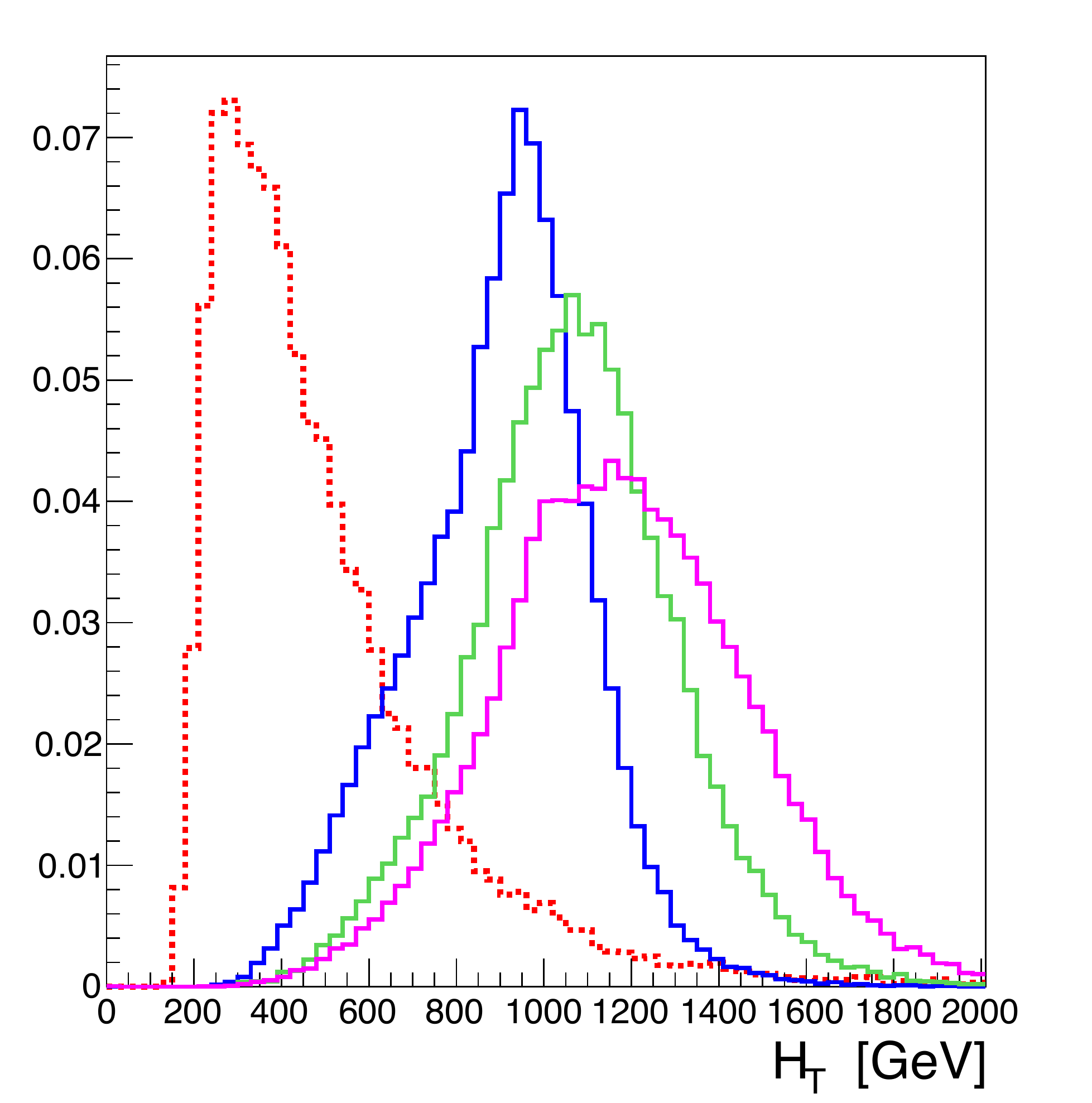}
\includegraphics[width=0.3\textwidth, height=0.28\textwidth, clip,angle=0]{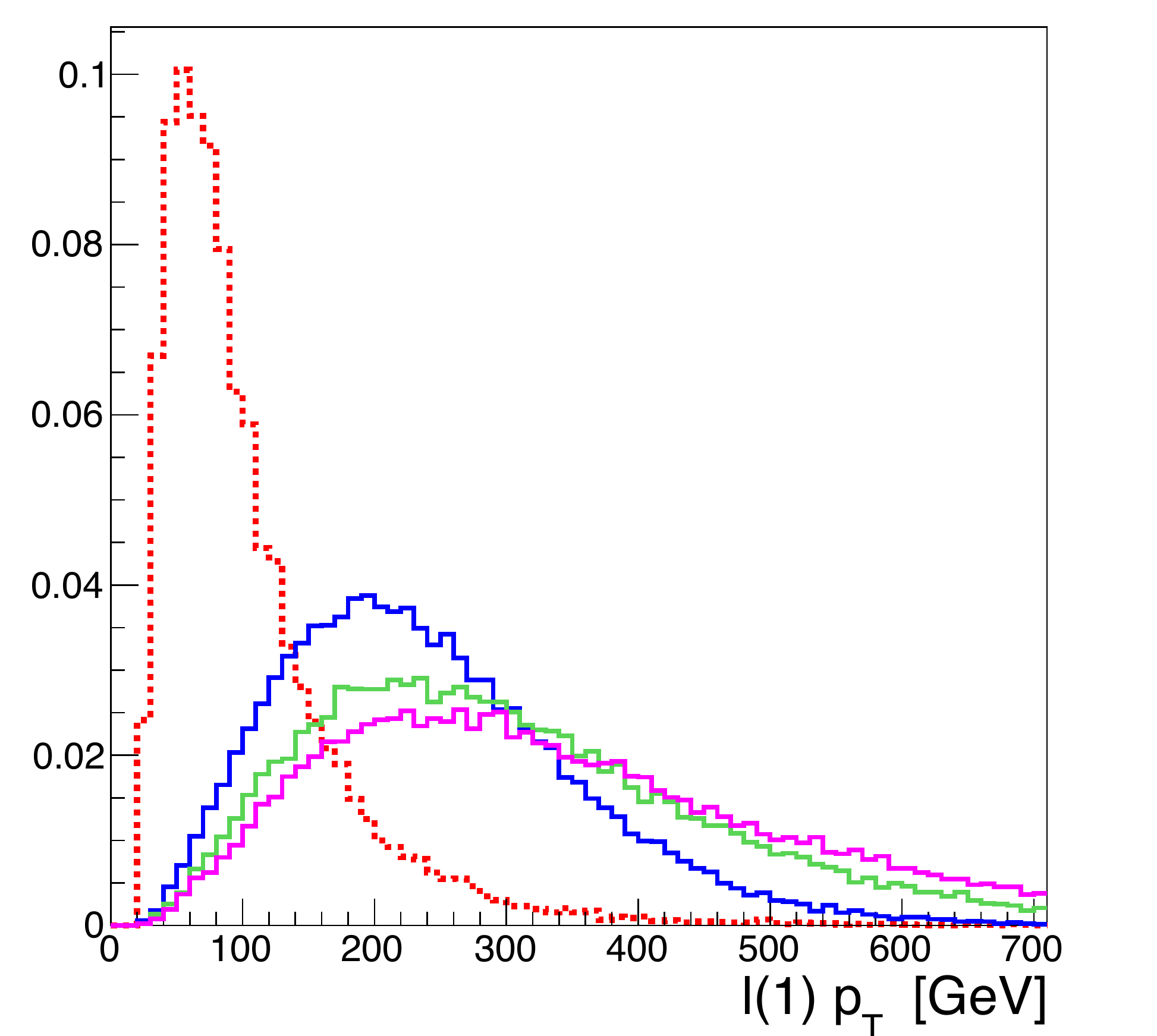}
\includegraphics[width=0.31\textwidth, height=0.305\textwidth,clip,angle=0]{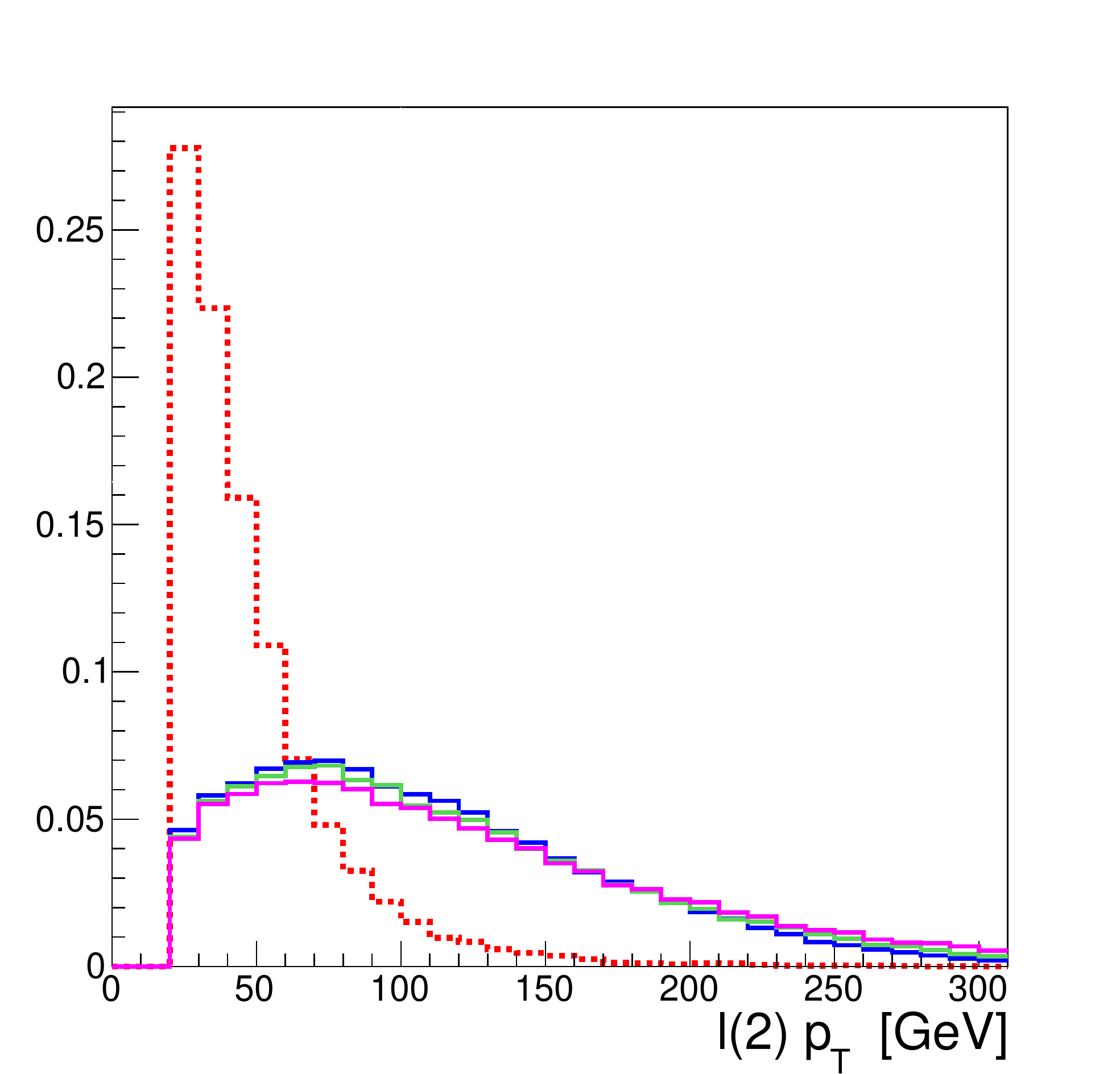}
\caption[]{
\label{fig:dist-cust}
\small
$S_T$, $H_T$ and $p_T$ normalized distributions for the signal $W^{' +}\to T_{5/3} \bar{T}_{2/3}$ at different $W^{'}$ masses (solid curves) and the total background (red dotted curve) at the 14 TeV LHC.
}
\end{center}
\end{figure}

Signal and background cross section after acceptance cuts are shown in the second column of Tab. \ref{tab:cutflow14TeV1}. The background is not large and can be reduced to a very low rate with a simple selection that exploits the characteristic energetic final state of the signal and mainly relies on $p_T$ cuts. Specifically, we apply cuts on the transverse momentum of the leading and second leading lepton, of the leading and second leading jet, a cut on $H_T$, defined as the scalar sum of the $p_T$ of all the final jets and leptons and a cut on $S_T$, $S_T\equiv H_T +\ptmiss$. Fig. \ref{fig:dist-cust} shows the $p_T$, $H_T$ and $S_T$ distributions for the signal at different $W^{'}$ masses and for the total background.
The values of the cuts we impose are shown in Tab. \ref{tab:cut}. For $m_{W^{'}} \geq 2.5$ TeV the selection is refined by strengthening the $H_T$ and $S_T$ cuts. We name CUT-1 the set of cuts with lower $H_T$ and $S_T$ cuts and CUT-2 the set with harder cuts.  \\

\begin{figure}[]
\begin{center}
\includegraphics[width=0.35\textwidth,clip,angle=0]{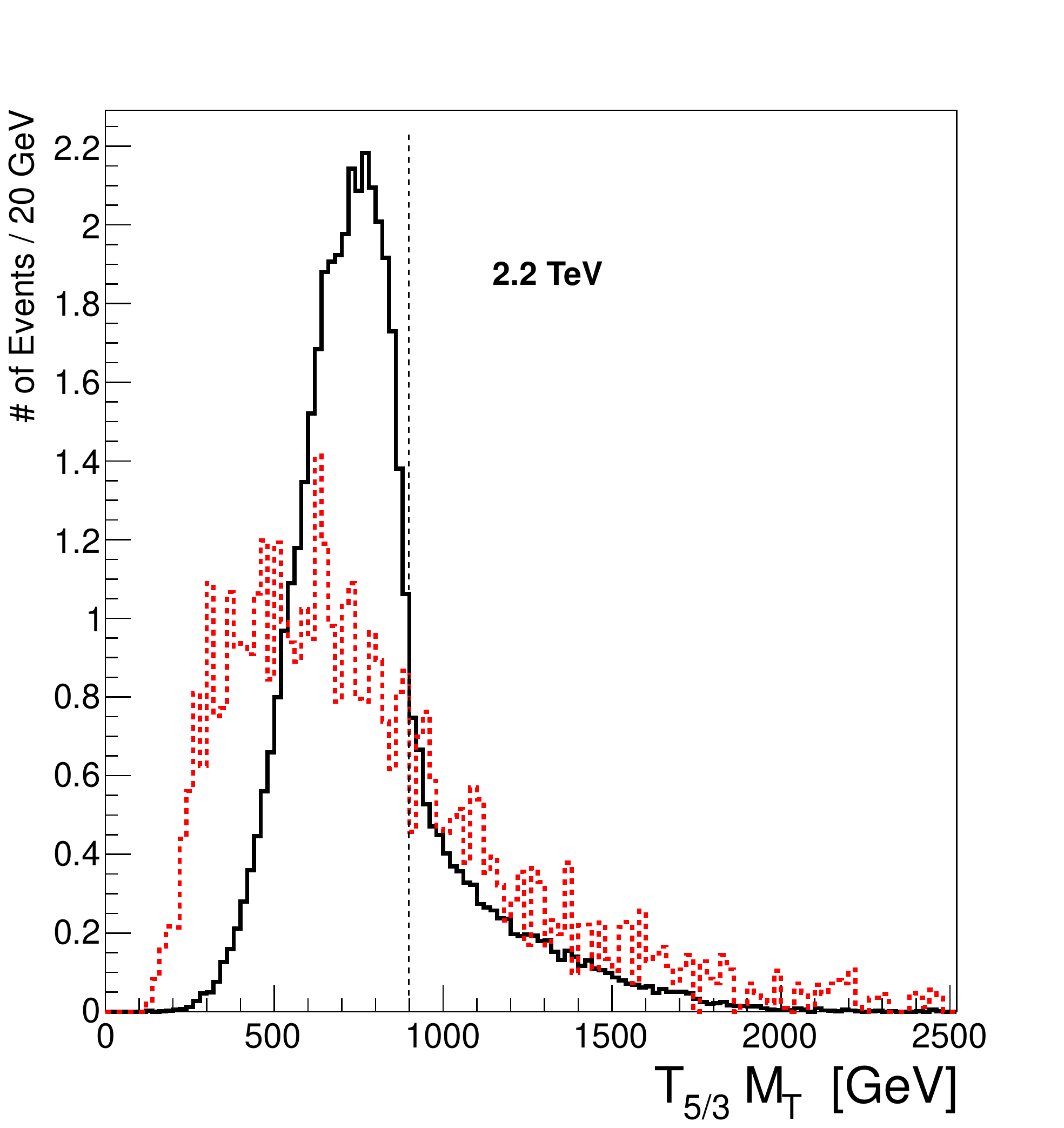}
\includegraphics[width=0.39\textwidth,clip,angle=0]{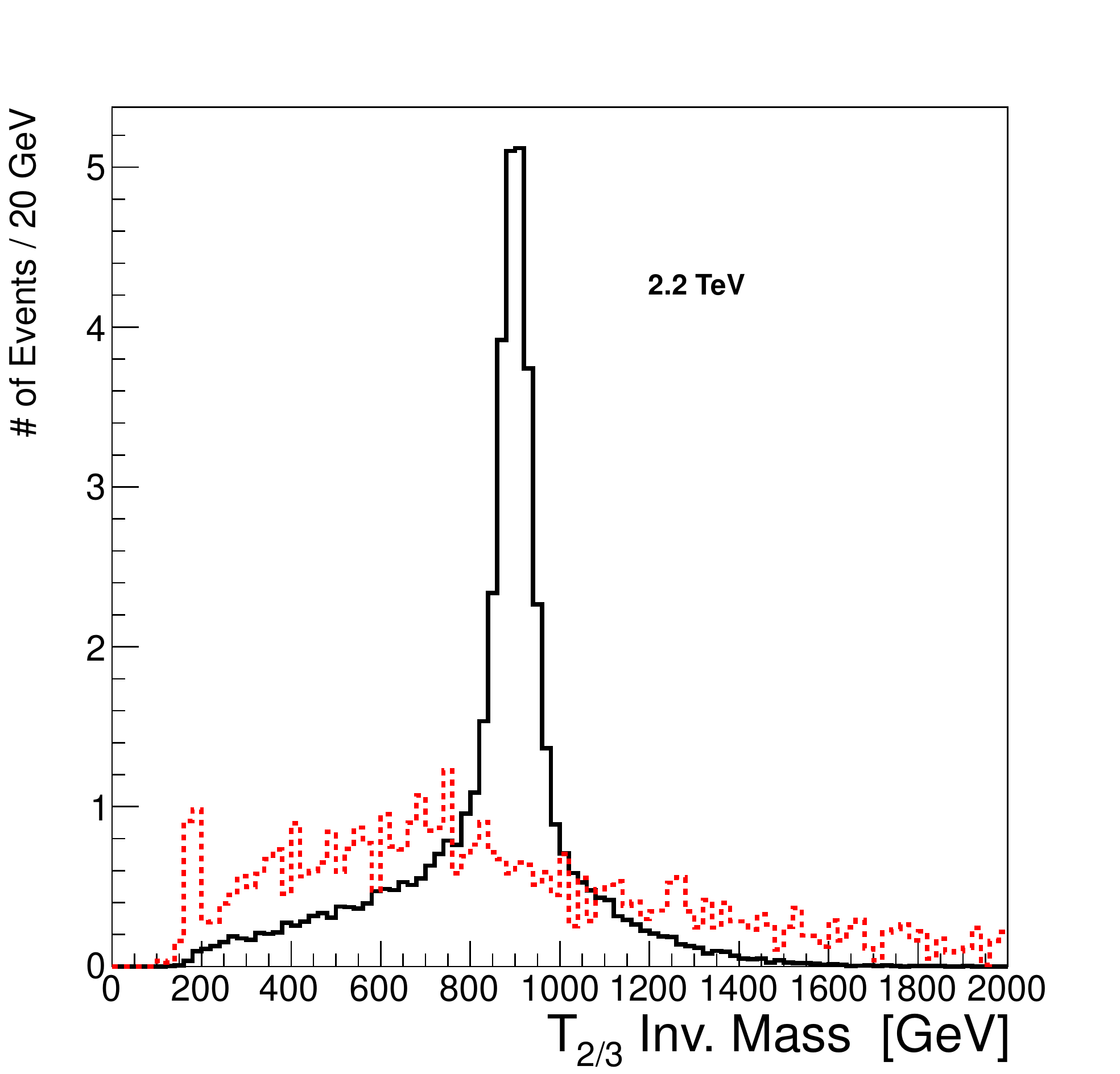}
\caption[]{
\label{fig:cust-mass}
\small
Transverse mass of the reconstructed $T_{5/3}$ (left plot) and invariant mass of the reconstructed $T_{2/3}$ (right plot) for the signal $W^{'}\to T_{5/3}T_{2/3}$ with $m_{W^{'}}=2.2$ TeV ($m_C=0.9$ TeV) and for the total background (red dotted curves), after the main CUT-1 selection in Tab. \ref{tab:cut} at the 14 TeV LHC with 100 fb$^{-1}$. The dotted vertical line in the left panel indicates the position of the expected Jacobian peak.
}
\end{center}
\end{figure}

After the main selection we proceed with the reconstruction of the custodian heavy resonances. This is an important step to further reduce the background and also to estimate the value of the $W^{'}$ mass, in the hopefully case of a discovery. Due to the presence of two neutrinos in the final state, indeed, the $W^{'}$ cannot be completely reconstructed, but one can infer the $W^{'}$ mass value by analyzing the total transverse mass and also the $p_T$ distributions of the $W^{'}$ decay products, the $T_{5/3}$ and $T_{2/3}$ heavy fermions. Obviously the reconstruction of the custodian resonances is also crucial to claim a discovery for the custodians themselves. We apply a simple reconstruction procedure. We select as the jet associated with the $T_{5/3}$ decay (Fig. \ref{fig:feyn-cust}) the one among all of the final reconstructed jets with the lowest $\Delta R$ separation (but greater than 0.4) from the second leading lepton. Looking at the $T_{5/3}$ decay products, $T_{5/3}\to (W\to l\nu)(t\to l\nu j)$, we find indeed that the lepton from the top is typically softer than the lepton from the $W$ and that this second leading lepton tends to be close to the jet, which also comes from the top decay.  Having tagged the jet from $T_{5/3}$ we can calculate the $T_{5/3}$ total transverse mass and we can reconstruct the $T_{2/3}$ completely, by considering as its decay products all of the final reconstructed jets with the exclusion of the tagged jet from $T_{5/3}$. The resulting $T_{5/3}$ transverse mass and the $T_{2/3}$ invariant mass are shown in Fig. \ref{fig:cust-mass} for the signal with $m_{W^{'}}=2.2$ TeV and the total background, after the main selection, at the 14 TeV LHC with 100 fb$^{-1}$. Once having reconstructed the custodians we complete our selection by applying a cut on the $T_{2/3}$ invariant mass and a (mild) cut on the transverse mass of the reconstructed $T_{5/3}$: \footnote{We define the transverse mass as 
\[
M^2_T= \Bigl ( \sqrt{ |\vec{p}_T|^2 +M^2 } + |\vec{\ptmiss}|  \Bigr )^2  -| \vec{p}_T +\vec{\ptmiss} |^2 \ ,
\]
where $\vec{p}_T$ ($M$) is the transverse momentum (the invariant mass) of the system $\{ l^+ l^+ j \}$, made up of the $T_{5/3}$ decay products, in the case of the $T_{5/3}$ transverse mass, and of the system $\{ l^+ l^+ +n\ \text{jets} \} $, in the case of the total transverse mass. 
}

\begin{equation}\label{eq:cust-cut}
M_{T_{2/3}}\in [0.8, 1.0] \, \text{TeV} \qquad  M_T (T_{5/3}) > 400 \, \text{GeV}  \ .
\end{equation}

\begin{table}
\begin{center}
{\small
\begin{tabular}{|c|c||c|c||c|c||}
\hline
& & & & &  \\
 & & & &  & \\[-0.6cm]
 \multicolumn{1}{|c|}{\textsf{Same-Sign Dilep}} & acceptance & CUT-1 & FINAL-1 & CUT-2 & FINAL-2 \\[0.2cm]
 \hline
 & & & & &  \\
  & & & & & \\[-0.4cm]
 $m_{W^{'}}=1.9$ TeV            & 0.82 &  \bf{0.66} & \bf{0.39} & 0.43 & 0.25  \\[0.25cm]
 $m_{W^{'}}=2.2$ TeV            & 0.52 &  \bf{0.45} & \bf{0.27} & 0.37 & 0.22  \\[0.25cm]
  $m_{W^{'}}=2.5$ TeV            & 0.29 &  0.24 & 0.15 & \bf 0.21 & \bf{0.14}  \\[0.25cm]
   $m_{W^{'}}=3.0$ TeV            & 0.11 &  0.076 & 0.052 & \bf 0.070 & \bf{0.048}  \\[0.25cm]
   $m_{W^{'}}=3.5$ TeV            & 0.041 &  0.036 & 0.025 & \bf 0.033 & \bf{0.024}  \\[0.25cm]
\hline
 & & & & &  \\
  & & & & & \\[-0.4cm]
  $W^{+}t\bar t$            & 4.1 &  0.20 & 0.021 & 0.06 & 0.009  \\[0.17cm]
  $W^{+}W^{+}$            & 1.5 &  0.20 & 0.024 & 0.12 & 0.017  \\[0.17cm]
  $W^{+}W^{+}W^{-}$            & 0.6 &  0.06 & 0.006 & 0.03 & 0.002  \\[0.25cm]
  Total                  & & & & & \\
  background      &  6.2 & 0.46  & 0.051  & 0.21 & 0.028  \\[0.15cm]
\hline
\end{tabular}
}
\caption{
\label{tab:cutflow14TeV1}
\small 
Cross sections, in fb, at $\sqrt{s}=14\,$TeV for the signal $W^{'}\to T_{5/3} \bar T_{2/3}$  (with $\cot\theta_2=3$) and the main backgrounds after imposing the acceptance cuts of 
eq.(\ref{eq:evsel-cust}) (second column), after the main cuts of Tab. \ref{tab:cut} (third column for CUT-1 and fifth column for the harder CUT-2 selection) and after the further restriction on the reconstructed $T_{2/3}$ invariant mass  and on the reconstructed $T_{5/3}$ transverse mass in eq. (\ref{eq:cust-cut}) (fourth column after CUT-1, sixth column after CUT-2). For $m_{W^{'}} \geq 2.5$ TeV we get a higher significance by applying the harder CUT-2 selection.  
}
\end{center}
\end{table}

\begin{table}
\begin{center}
{\small
\begin{tabular}{c| c c}
& CUT-1 & CUT-2 \\[0.15cm]
\hline
& &   \\
 & &  \\[-0.6cm]
$p_T$ $l (1)$ & \multicolumn{2}{c}{90} \\[0.15cm]
$p_T$ $l (2)$ & \multicolumn{2}{c}{30}  \\[0.15cm]
$p_T$ $j (1)$ & \multicolumn{2}{c}{160}  \\[0.15cm]
$p_T$ $j (2)$ & \multicolumn{2}{c}{100}  \\[0.15cm]
$H_T$ & 550 & 700  \\[0.15cm]
$S_T$ & 1100 & 1400  \\[0.15cm]
\end{tabular}
}
\caption{
\label{tab:cut}
\small 
Main selection cuts, in GeV. For $m_{W^{'}} \geq 2.5$ TeV the selection is refined by strengthening the $H_T$ and $S_T$ cuts (CUT-2). 
}
\end{center}
\end{table}

Table \ref{tab:cutflow14TeV1} shows the values of the cross section for the signal $W^{'}\to T_{5/3} \bar T_{2/3}$  (with $\cot\theta_2=3$) and the background after each step of the analysis: the acceptance cuts, the main CUT-1 and CUT-2 selection and after the further restriction on the reconstructed $T_{2/3}$ invariant mass and on the reconstructed $T_{5/3}$ transverse mass. For $m_{W^{'}} \geq (<) \, 2.5$ TeV the significance is higher by applying the harder CUT-2 (looser CUT-1) selection. \\

\begin{figure}[]
\begin{center}
\includegraphics[width=0.35\textwidth,clip,angle=0]{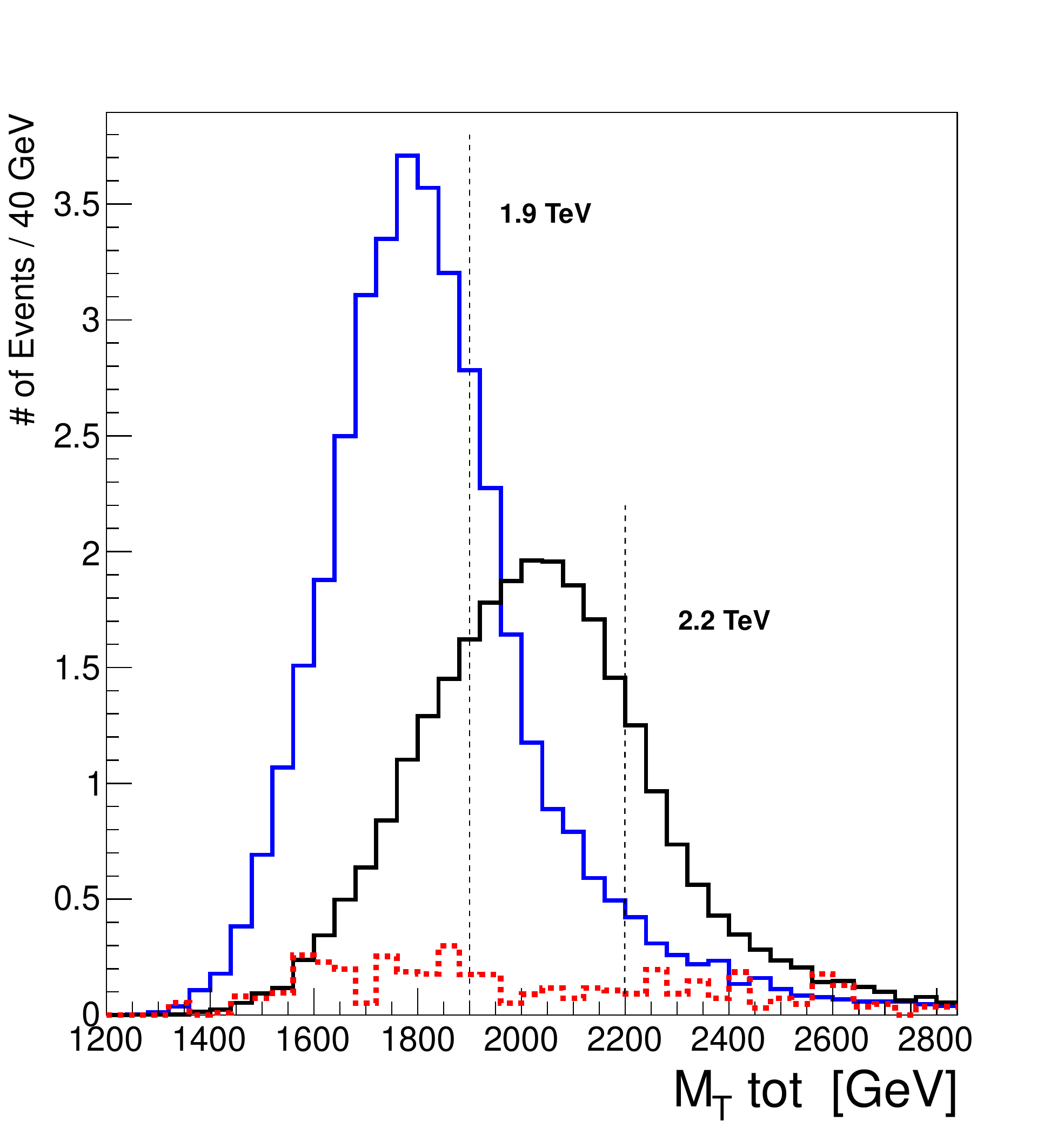}
\includegraphics[width=0.365\textwidth,clip,angle=0]{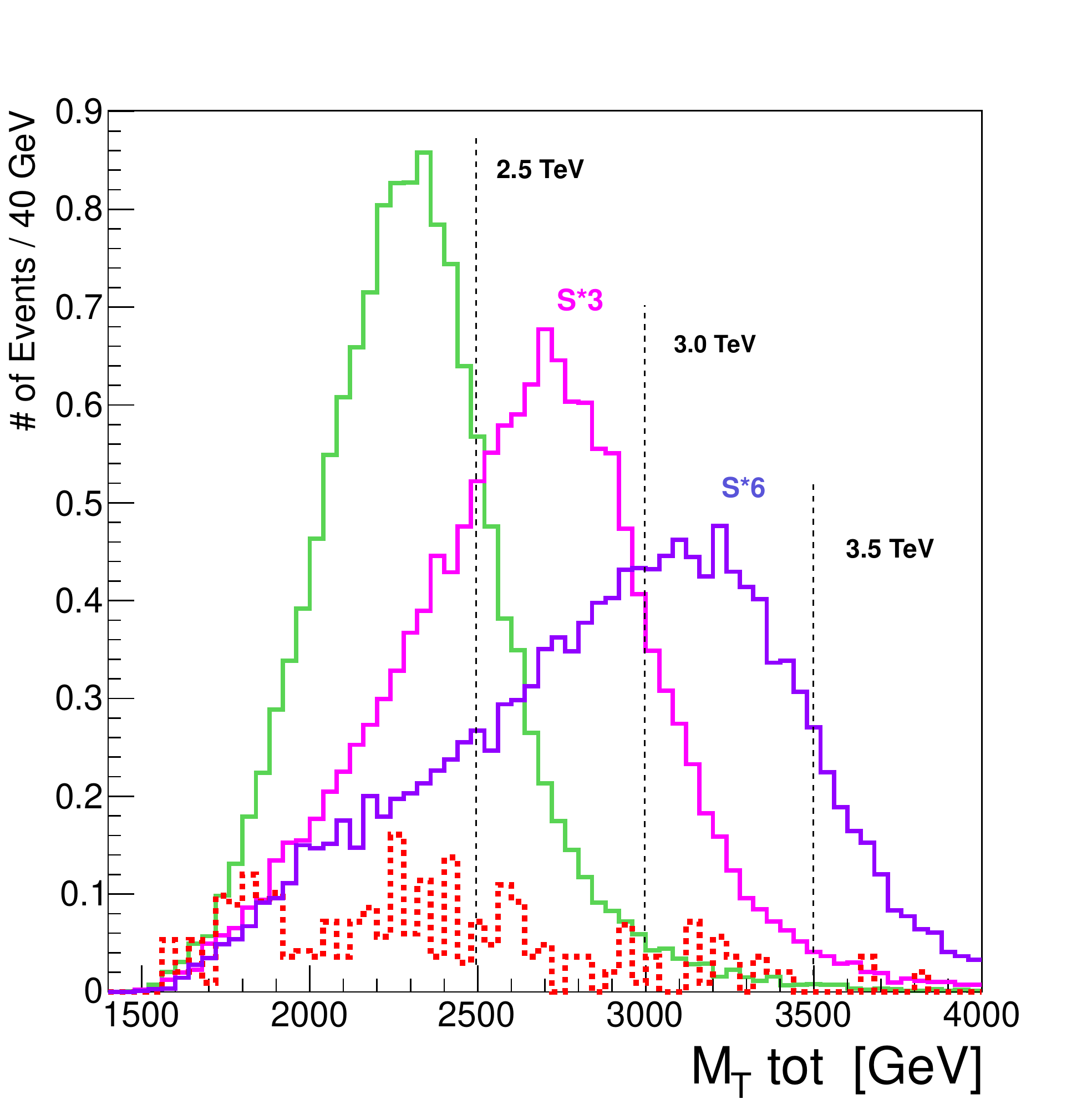}
\caption[]{
\label{fig:Mass-fin}
\small
Total transverse mass for the background (red dotted curve) and the signal $W^{'}\to T_{5/3}T_{2/3}$ (solid curves) 
at different $W^{'}$ masses at the 14 TeV LHC with 100 fb$^{-1}$, after the complete selection. We have applied the harder CUT-2 selection for $m_{W^{'}}\geq 2.5$ TeV (right plot). The dotted vertical lines show the position of the expected Jacobian peaks.
}
\end{center}
\end{figure}

\begin{figure}[]
\begin{center}
\includegraphics[width=0.36\textwidth,clip,angle=0]{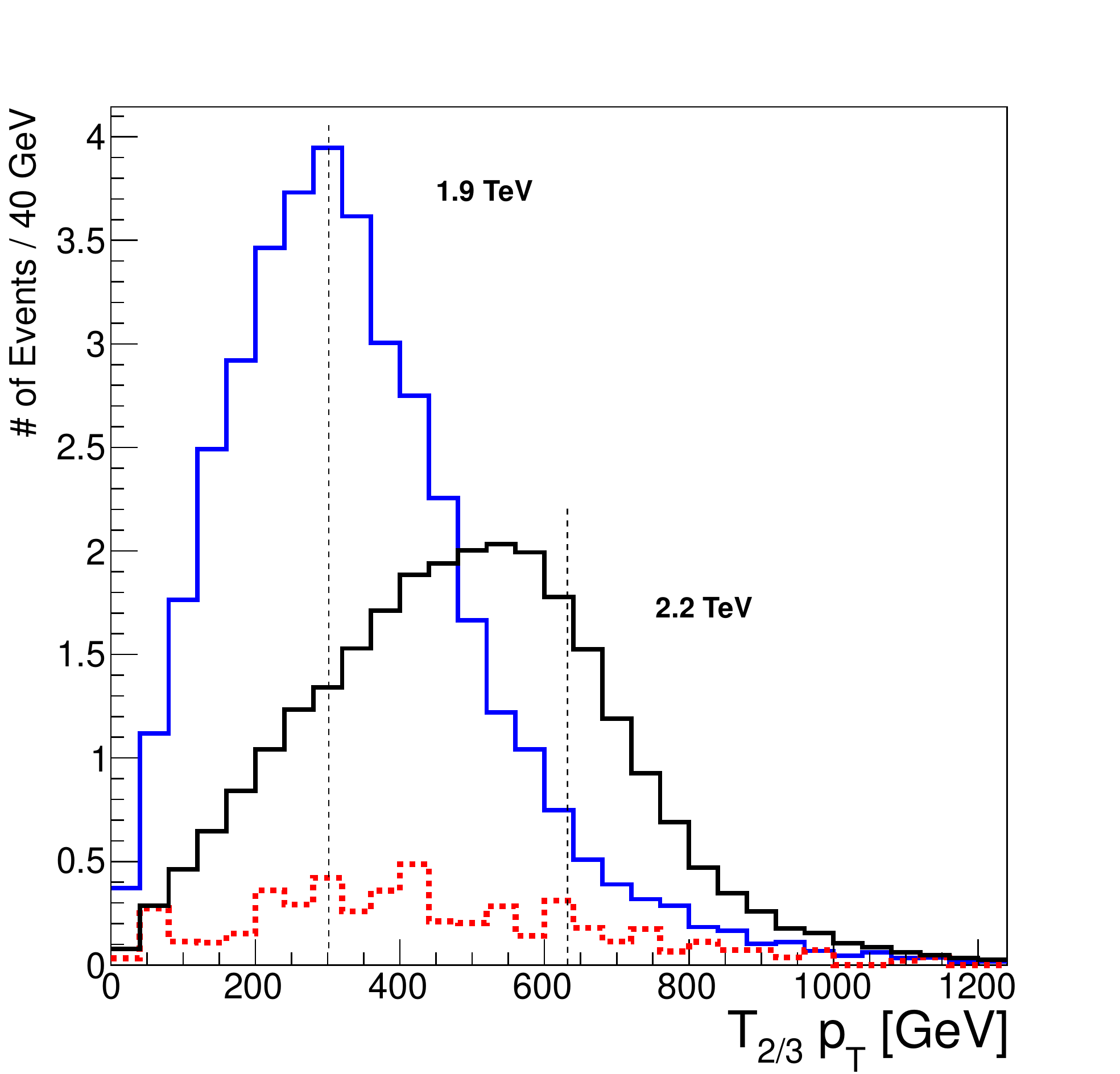}
\includegraphics[width=0.34\textwidth,clip,angle=0]{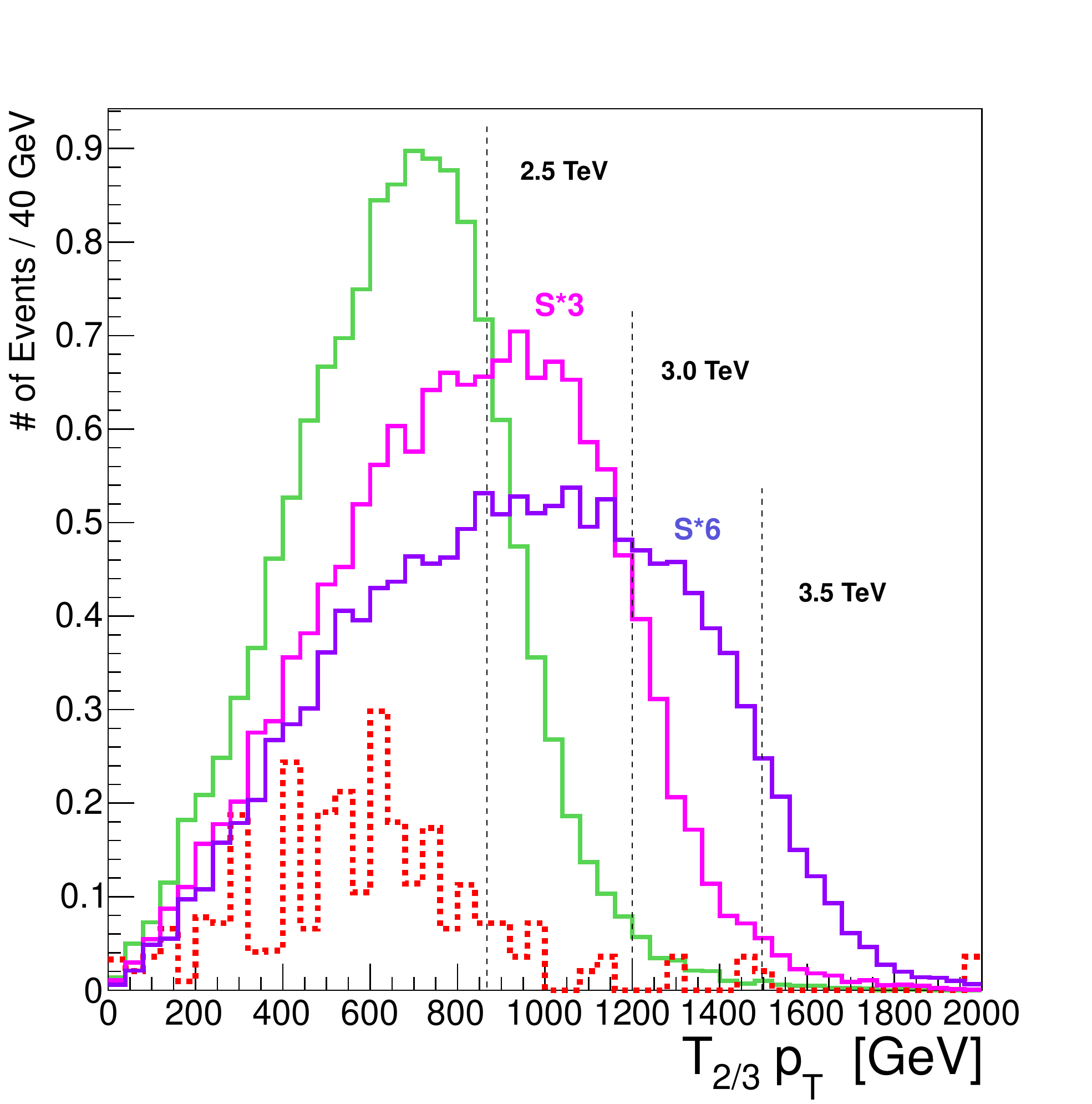}
\includegraphics[width=0.36\textwidth,clip,angle=0]{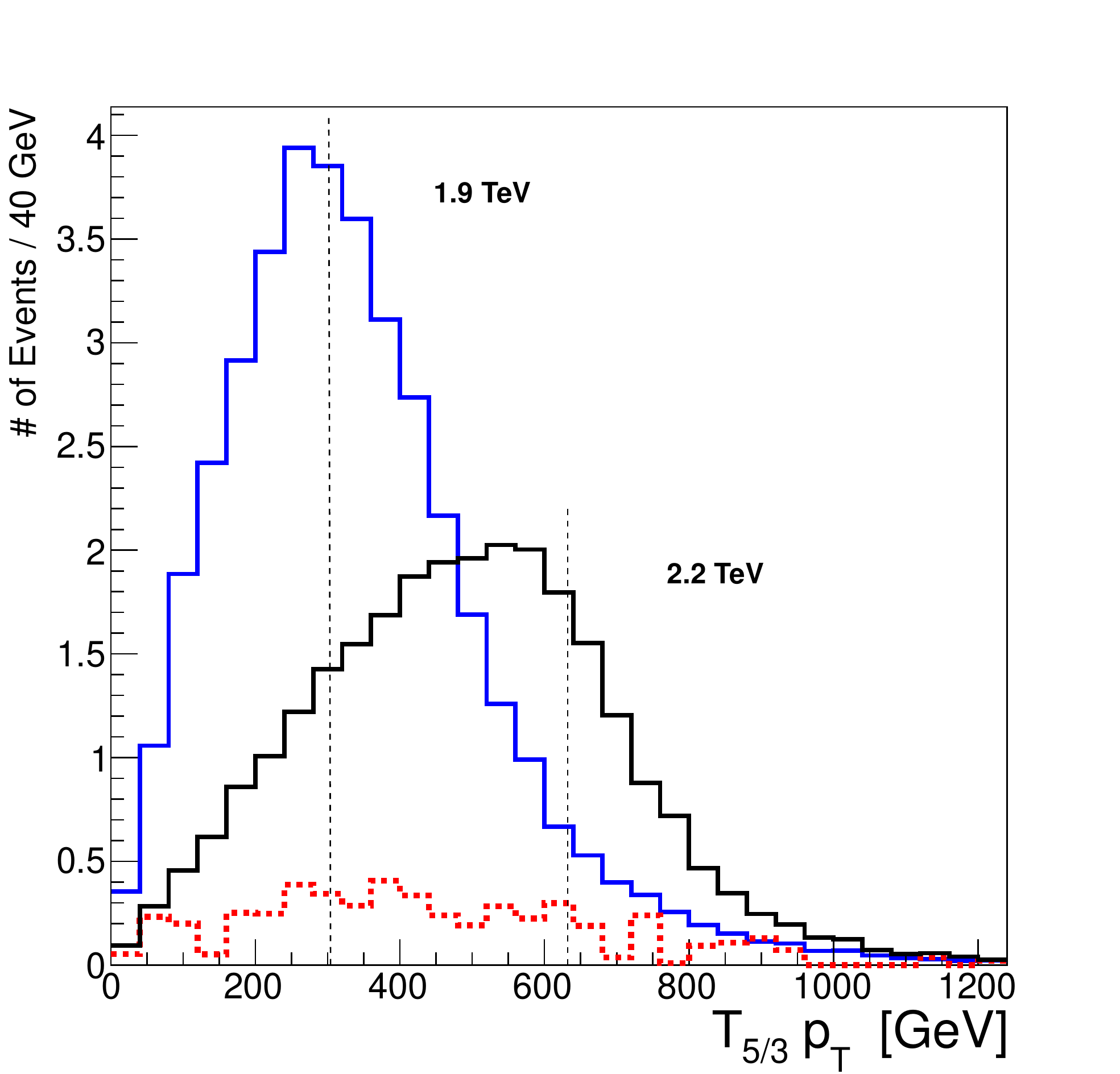}
\includegraphics[width=0.34\textwidth,clip,angle=0]{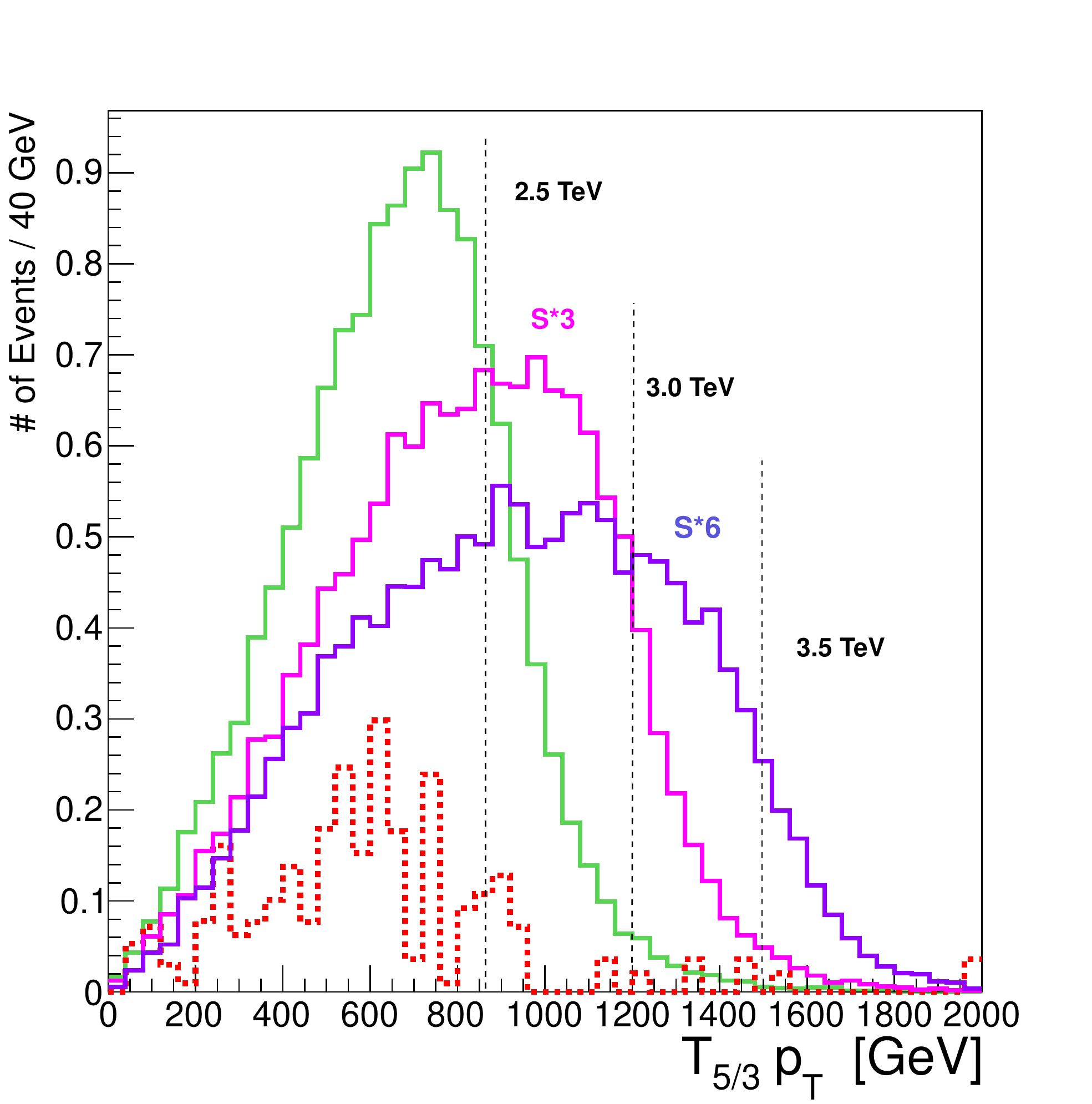}
\caption[]{
\label{fig:pt-cust}
\small
Reconstructed $T_{2/3}$ (upper plots)  and $T_{5/3}$ (lower plots) $p_T$ distributions for the background (red dotted curve) and the signal $W^{'}\to T_{5/3}T_{2/3}$ (solid curves) at different $W^{'}$ masses at the 14 TeV LHC with 100 fb$^{-1}$, after the complete selection. We have applied the harder CUT-2 selection for $m_{W^{'}}\geq 2.5$ TeV (right plots). The dotted vertical lines show the position of the expected Jacobian edges.
}
\end{center}
\end{figure}

Fig. \ref{fig:Mass-fin} shows the total transverse mass for signal and background at the 14 TeV LHC with 100 fb$^{-1}$, after the complete selection. Fig. \ref{fig:pt-cust} shows the $p_T$ distributions of the reconstructed custodians, which exhibit a Jacobian edge near the theoretically expected value of $\sqrt{m^2_{W^{'}}/4-m^2_C}$. The Jacobian edge shape of the distributions is smeared by the following effects: finite detector resolution, non-perfect reconstruction of the custodian resonances and non-zero $W^{'}$ transverse momenta. As already pointed out, one could infer the $W^{'}$ mass value by analyzing these total transverse mass and custodian $p_T$ distributions.\\

\begin{figure}[]
\begin{center}
\includegraphics[width=0.65\textwidth,clip,angle=0]{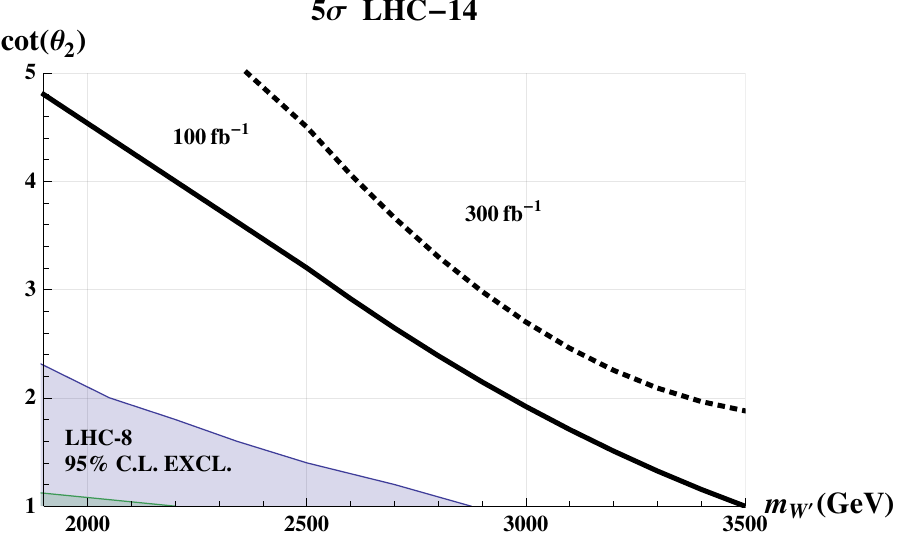}
\caption[]{
\label{fig:reach-cust}
\small
$pp \to \bf W^{'}\to T_{5/3} T_{2/3}$ to same-sign dilepton discovery reach on the $(m_{W'}, \cot\theta_2)$ plane at the 14 TeV LHC with 100 fb$^{-1}$  (solid curve)  and 300 fb$^{-1}$ (dotted curve). We set $s_L=0.9$ and $m_C=0.9$ TeV. We also show the region excluded by current LHC-8 analyses (derived in Sec. \ref{sec:limits}).
}
\end{center}
\end{figure}

Tab. \ref{tab:14tev-fin-cust} shows the results of the analysis after the complete selection for $\cot\theta_2=3$. We list the signal-over-background ratio, the discovery significance with 100 fb$^{-1}$ and the minimal integrated luminosity required for a 5 $\sigma$ discovery at the LHC with $\sqrt{s}=14$ TeV. As in the previous analysis of the $Tb$ channel, we will infer from these results the LHC-14 discovery reach on the full $(m_{W^{'}},\cot\theta_2)$ plane, by considering a scaling with $\cot\theta_2$ of the $W^{'}$ production cross section and of the $W^{'}\to T_{5/3}T_{2/3}$ branching ratio. We will also include the corrections from the $W^{'}-W^{'}_R$ electroweak mixing, which however are found to be small in the region at high $W^{'}$ mass (Sec. \ref{sec:app}). The result of our extrapolation is shown in Fig. \ref{fig:reach-cust}, where we plot the 5$\sigma$ discovery reach of the custodian channel at the LHC with 100 fb$^{-1}$ and 300 fb$^{-1}$. We also show the quite small portion of the parameter space, in the weakly-coupled region at small $\cot\theta_2$ values, excluded by the LHC-8 analyses of the $l\nu$ mode. We see that the custodian channel is a very promising signature. By analyzing this channel the 14 TeV LHC has the possibility to extensively test the high $W^{'}$ mass region. For a $W^{'}$ of about 2 TeV it is possible to have a discovery even in the more strongly-coupled scenarios, with $\cot\theta_2$ up to $\sim 5$. In the intermediate strongly-coupled regime with $\cot\theta_2\simeq 3$ we could discover a $W^{'}$ with masses up to 2.6 TeV with 100 fb$^{-1}$ and up to 2.9 TeV with 300 fb$^{-1}$.  \\

We conclude this section by pointing out that the analysis we have performed can be improved in several ways, for example by requiring a top tagging or, considering that the top quarks from the custodians have a fixed right-handed chirality \cite{Vignaroli:2012si}, by looking at angular distributions \cite{Berger:2011xk}.

\begin{table}
\begin{center}
{\small
\begin{tabular}{|c|c|c|c|c|c|}
\hline 
 & & & & & \\
 & & & & & \\[-0.6cm]
 \textsf{$W^{' +}\to T_{5/3}\bar{T}_{2/3}$} & $m_{W^{'}}=1.9$ TeV & $m_{W^{'}}=2.2$ TeV & $m_{W^{'}}=2.5$ TeV & $m_{W^{'}}=3.0$ TeV & $m_{W^{'}}=3.5$ TeV \\  [0.2cm]
 \hline
  & & & & & \\[-0.6cm]
  & & & & & \\
  $S/B$  & 7.6 & 5.3 & 4.9 &  1.7 & 0.9        \\[0.2cm]
  sign (100 fb$^{-1}$)            & 11 & 7.9 & 5.6 & 2.2 & 1.1 \\[0.2cm]
  5$\sigma$ Int. Lum.        & 24 fb$^{-1}$ & 40 fb$^{-1}$ & 80 fb$^{-1}$ &  450 fb$^{-1}$ & 1500 fb$^{-1}$    \\[0.2cm]
\hline
\end{tabular}
}
\caption{
\label{tab:14tev-fin-cust}
\small 
signal-over-background ratio, discovery significance with 100 fb$^{-1}$, and minimal integrated luminosity required for a 5 $\sigma$ discovery at the LHC with $\sqrt{s}=14$ TeV for the signal $W^{' +}\to T_{5/3}\bar{T}_{2/3}$ to same-sign dileptons (with $\cot\theta_2=3$) . 
}
\end{center}
\end{table}

\section{Conclusions}
\label{sec:conclusion}

We have analyzed the phenomenology of a $W^{'}$ vector boson from a new composite/warped extra dimensional sector in the scenario, favored by both naturalness argument and indications from electroweak precision tests, where the $W^{'}$ mass is above the threshold for $W^{'}$ decays into one or two vector-like top partners, the lightest of which we have assumed to be slightly below 1 TeV.\\
We have extracted bounds on the $W^{'}$ mass and couplings from the current LHC-8 analyses (Fig. \ref{fig:limits}), finding that the searches in the `standard' decay modes, $l\nu$, $WZ$, $jj$, $tb$ can exclude a significant parameter region at lower $W^{'}$ masses, but have basically no-sensitivity to the higher mass region $m_{W^{'}}\gtrsim 2$ TeV. \\
We have shown how this heavier $W^{'}$ region, favored by the constraints on the $S$ parameter, can be extensively explored at the 14 TeV LHC, by analyzing the $W^{'}$ decay into a pair of custodian top partners. We have studied in particular the $W^{'} \to T_{5/3}T_{2/3}$ channel in the same-sign dilepton final state, finding that the search could give a discovery at the 14 TeV LHC  for a $W^{'}$ with a mass up to  2.6 TeV with 100 fb$^{-1}$ and up to 2.9 TeV with 300 fb$^{-1}$, in an intermediate strongly-coupled scenario with $\cot\theta_2=3$. The discovery reach on the full parameter space is shown in Fig. \ref{fig:reach-cust}.\\
We have also studied the $W^{'}$ decay into a top-prime plus a SM bottom quark. This heavy-light signature is powerful to test the intermediate $W^{'}$ mass region for not-too-large top degrees of compositeness. We have analyzed the heavy-light channel in the semileptonic final state at the 8 TeV and 14 TeV LHC. With approximately 20 fb$^{-1}$ we find that the channel could slightly extend the limits on $W^{'}$ mass and couplings resulting from the current analyses at the 8 TeV LHC. At the 14 TeV LHC the heavy-light channel could test the intermediate $W^{'}$ mass region widely and a discovery could occur even in the more strongly-coupled regime. The exclusion/discovery potential of the channel is shown in Fig. \ref{fig:reach-Hl-8tev} for the 8 TeV LHC and in Fig. \ref{fig:reach-Hl-14tev} for the 14 TeV LHC.\\
We conclude by pointing out that the new $W^{'}$ to vector-like quarks signals are promising channels for the discovery of the top partners as well and that the phenomenology here delineated for a $W^{'}$, where the decays to top partners are the dominant modes, is similar for a $Z^{'}$ and for a heavy gluon $G^{'}$ too. In particular, the high sensitivity of the $G^{'}$ heavy-light decay channels has been already stressed in \cite{Bini:2011zb}. For a $G^{'}$, on the other hand, the custodian channel is probably less powerful than for a $W^{'}$, due to a larger total decay width.

\section*{Acknowledgments}

I thank Sekhar Chivukula and C.-P. Yuan for useful discussions and comments on the manuscript. This work is supported by the U.S. National Science Foundation under Grant No. PHY-0854889.

\appendix
\section{$W^{'}-W^{'}_R$ mixing and $W^{'}_R$ phenomenology}
\label{sec:app}

We will briefly discuss in this section the phenomenology of a $W^{'}_R$ from a $SU(2)_R$ composite triplet (singlet under $SU(2)_L$) and the effect on the $W^{'}$ phenomenology of its electroweak mixing with the $W^{'}$.

The mass matrix that determines the $W^{'}-W^{'}_R$ mixing after the EWSB is given, in the $(W^{'}, W^{'}_{R})$ basis, by \cite{Contino:2006nn}:

\begin{equation}\label{eq:LRmix}
m^2_{\pm}= \frac{v^2}{2} \begin{bmatrix} & \frac{2 m^{2}_{*2}}{v^2 \cos^2\theta_2} +g^2_2\cot^2\theta_2 & -g_1 g_2 \frac{\cot\theta_2}{\sin\theta_1} \,  \\ & -g_1 g_2 \frac{\cot\theta_2}{\sin\theta_1} & \frac{2 m^{2}_{*1}}{v^2} +g^2_2\frac{1}{\sin^2\theta_2} \, \end{bmatrix} \normalsize \, ,
\end{equation}
where $v=174$ GeV, $m_{*2}, m_{*1}$ are the bare (before any mixings) $W^{'}$ and $W^{'}_R$ masses respectively and $g_{1}=e/\cos\theta_W,\, g_2=e/\sin\theta_W$. $\theta_1$ is the elementary-composite mixing angle in the hypercharge sector:
 
\begin{equation}\label{eq:ct1}
\cot\theta_1=\frac{g^{*}_1}{g^{el}_{1}} \qquad g_1=g^{el}_{1}\cos\theta_1=g^{*}_{1}\sin\theta_1 \ .
\end{equation}   
\noindent
In what follows we will assume $m_{*1}= m_{*2}$ and $g^{*}_{1}=g^{*}_{2}$.
In (\ref{eq:LRmix}) we have neglected mixing terms for the SM $W$, which are very small for $W^{'}$, $W^{'}_R$ bosons above 1 TeV.  \\
As an effect of the mixing (\ref{eq:LRmix}), the physical $\bf W^{'}$ after the EWSB becomes a superposition of the before-EWSB $W^{'}$ and $W^{'}_R$ states:
 \begin{equation}\label{eq:Wp-mix}
\mathbf{W^{'}}  = c_{\xi} W^{'} - s_{\xi} W^{'}_R 
 \end{equation}
and the $\bf W^{'}_R$ is given by the orthogonal combination
 \begin{equation}
\mathbf{W^{'}_R}  = c_{\xi} W^{'}_R + s_{\xi} W^{'} \ .
 \end{equation}
We can notice that the $W^{'}$ rates in (\ref{eq:decays}) will be now multiplied by $c^{2}_{\xi}$ and new decay modes, coming from the $W^{'}_R$ interactions, will appear with a direct proportionality of their rates to $s^2_{\xi}$:

\begin{align}\label{eq:decay-mix}
\begin{split}
 \Gamma[\mathbf{W^{'+}} \to W^{+}Z]=\Gamma[\mathbf{W^{'+}} \to W^{+}h] & = c^{2}_{\xi} \, \Gamma[W^{'+} \to W^{+}Z]+s^{2}_{\xi}\, \frac{g^2_1}{192 \pi}\frac{m_{W^{'}}}{\sin^2\theta_1}\\
 \Gamma[\mathbf{W^{'+}} \to \bar{f} f^{'} ] & =c^{2}_{\xi} \, \Gamma[W^{'+} \to \bar{f} f^{'} ] \\
 \Gamma[\mathbf{W^{'+}} \to T \bar{b}]=\Gamma[\mathbf{W^{'+}} \to t\bar{B}] & = c^{2}_{\xi} \, \Gamma[W^{'+} \to T \bar{b}]\\
 \Gamma[\mathbf{W^{'+}} \to T_{5/3}\bar{T}_{2/3}] & = c^{2}_{\xi} \, \Gamma[W^{'+} \to T_{5/3}\bar{T}_{2/3}]\\
  \Gamma[\mathbf{W^{'+}} \to T\bar{B}] & = c^{2}_{\xi} \, \Gamma[W^{'+} \to T\bar{B}]\\
\Gamma[\mathbf{W^{'+}} \to T_{5/3}\bar{t}]\simeq\Gamma[\mathbf{W^{'+}} \to T_{2/3}\bar{b}] & = s^{2}_{\xi} \, \frac{g^2_1}{16 \pi}\frac{m_{W^{'}} }{\sin^2\theta_1} s^2_L \left(1-\frac{1}{2}\frac{m^2_C}{m^2_{W^{'}}}-\frac{1}{2}\frac{m^4_C}{m^4_{W^{'}}}  \right)\left(1-\frac{m^2_C}{m^2_{W^{'}}} \right)\\
\Gamma[\mathbf{W^{'+}} \to T_{5/3}\bar{T}]\simeq\Gamma[\mathbf{W^{'+}} \to T_{2/3}\bar{B}] & = s^{2}_{\xi} \, \frac{g^2_1}{16 \pi}\frac{m_{W^{'}} }{\sin^2\theta_1}\Biggl\{ \Biggr. \left(1+c^2_L\right) \left[ 1 - \frac{1}{2} \left( \frac{m^2_T}{m^2_{W^{'}}}+\frac{m^2_C}{m^2_{W^{'}}} \right) - \frac{1}{2} \left( \frac{m^2_T}{m^2_{W^{'}}}-\frac{m^2_C}{m^2_{W^{'}}} \right)^2   \right] \\
& +6 c_L\frac{m_C m_T}{m^2_{W^{'}}} \Biggl. \Biggr\} \sqrt{\left(1-\frac{m^2_T}{m^2_{W^{'}}} - \frac{m^2_C}{m^2_{W^{'}}}\right)^2- 4 \frac{m^2_T m^2_C}{m^4_{W^{'}}}}
\end{split}
\end{align}

\noindent
where $f$ indicates a generic SM fermion and $m_{W^{'}}$ is now the $W^{'}$ mass after the EWSB. This determines a modification in the $W^{'}$ branching ratios and a reduction, by a factor $c^{2}_{\xi}$, of the $W^{'}$ Drell-Yan production rate.\\

\begin{figure}[]
\begin{center}
\includegraphics[width=0.37\textwidth,clip,angle=0]{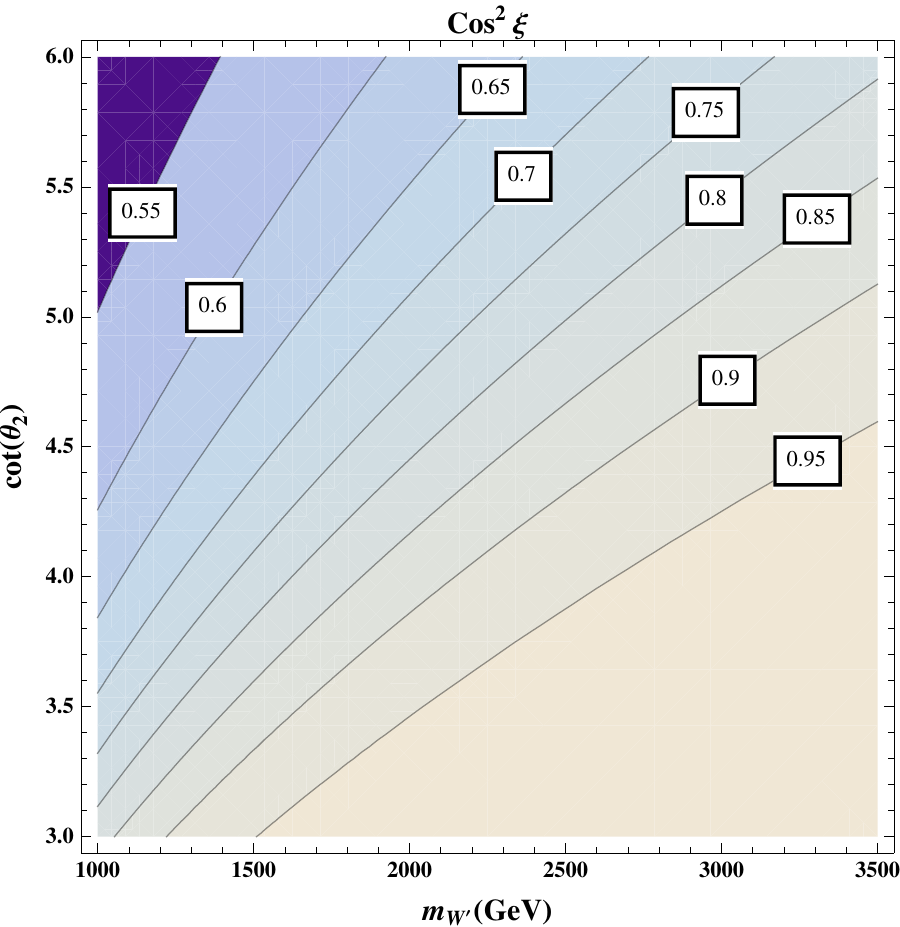}
\caption[]{
\label{fig:cosxi}
\small
value of the $c^2_{\xi}$ parameter in (\ref{eq:Wp-mix}) in the $(m_{W^{'}}, \cot\theta_2)$ plane. The Drell-Yan $W^{'}$ production rate gets a $c^2_{\xi}$ correction after the EWSB. 
}
\end{center}
\end{figure}

\begin{figure}[]
\begin{center}
\includegraphics[width=0.33\textwidth, clip,angle=0]{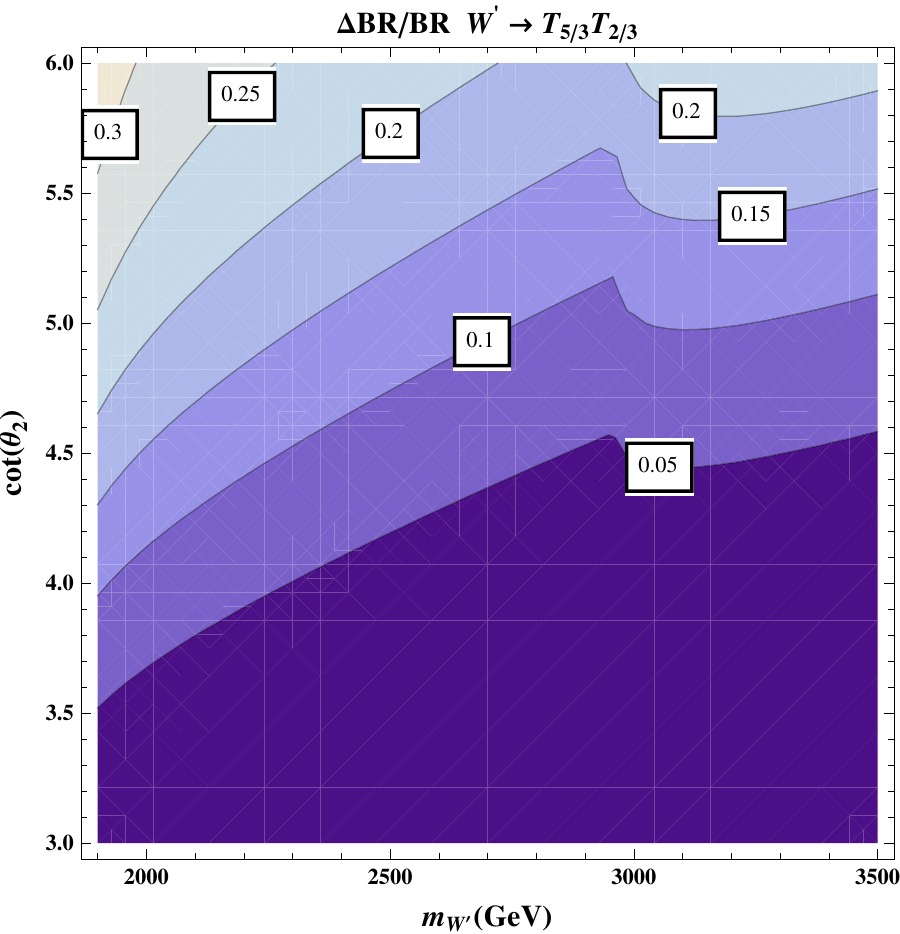} \ \ 
\includegraphics[width=0.33\textwidth,clip,angle=0]{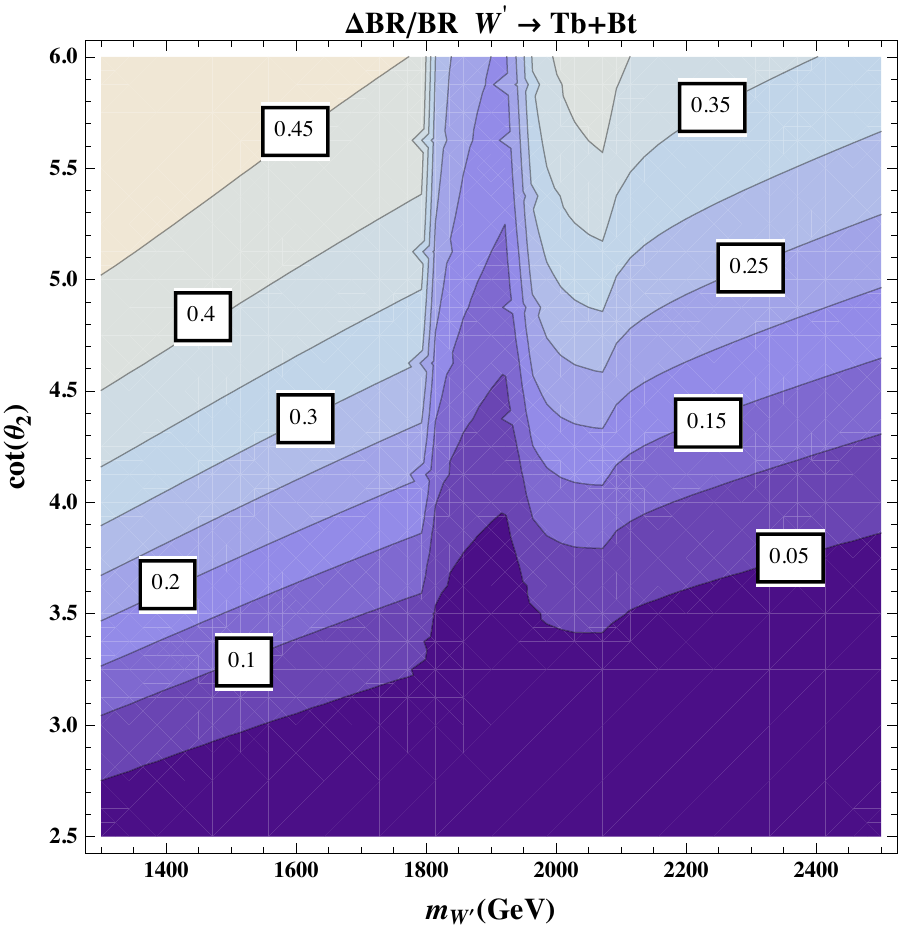}
\caption[]{
\label{fig:BR-mix}
\small
$\Delta BR/BR$ variation, after the EWSB, of the $W^{'}$ decay branching ratio into custodians for $s_L=0.9$ (left plot) and into $Tb$ for $s_L=0.5$ (right plot) in the $(m_{W^{'}}, \cot\theta_2)$ plane. 
}
\end{center}
\end{figure}

Fig. \ref{fig:cosxi} shows the value of $c^2_{\xi}$ in the $(m_{W^{'}}, \cot\theta_2)$ plane, which have been obtained by diagonalizing numerically the mixing matrix (\ref{eq:LRmix}). Fig. \ref{fig:BR-mix} shows the variation of the $W^{'}$ decay branching ratio into custodians for $s_L=0.9$ (left plot) and into $Tb$ for $s_L=0.5$ (right plot). We have fixed $m_C=0.9$ TeV, as in the analysis.

We see that the electroweak corrections are almost negligible in the more weakly-coupled regime, $\cot\theta_2 \lesssim 3$, but become significant in the intermediate mass region for larger $\cot\theta_2$ values, $\cot\theta_2 \gtrsim 4$. The resulting modification to the $W^{'}$ production cross section and decay branching ratios have been included in our estimate of the discovery/exclusion power of the different search channels. \footnote{We have not included corrections to the $W^{'}$ total decay width, which are typically small, below 5 percent.}  \\

\begin{figure}[]
\begin{center}
\includegraphics[width=0.47\textwidth,clip,angle=0]{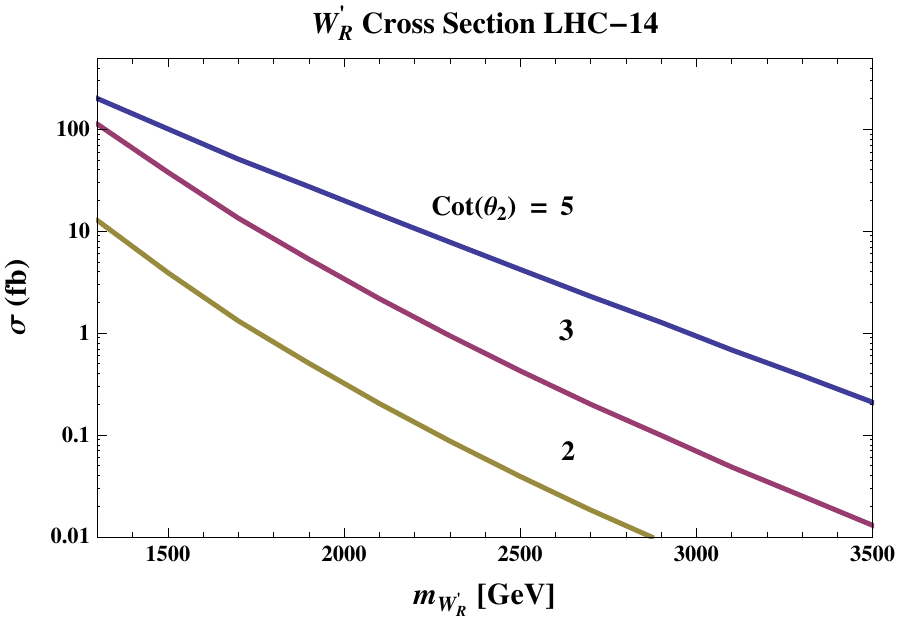}
\caption[]{
\label{fig:WRxsec}
\small
$W^{'}_R$ Drell-Yan production cross section at the 14 TeV LHC for different $\cot\theta_2$ values.
}
\end{center}
\end{figure}

We can also briefly discuss the $W^{'}_R$ phenomenology at the LHC. Similar studies have been performed in \cite{Agashe:2008jb, Grojean:2011vu}. \\
The $\bf W^{'}_R$ can interact with light quarks only through its $W^{'}$ component. Its Drell-Yan production is thus proportional to $s^{2}_{\xi}$ and has a larger rate for a larger mixing. We show in Fig. \ref{fig:WRxsec} the $ W^{'}_R$ Drell-Yan cross section at the LHC for different $\cot\theta_2$ values, that is for different size of the $W^{'}-W^{'}_R$ mixing. We see that $ W^{'}_R$ is produced at a much lower rate than $W^{'}$; nevertheless the events yield in the more strongly-coupled regime at $\cot\theta_2\sim 5,6$ could be probably sufficient for an observation at the 14 TeV LHC. This statement clearly needs further exploration.\\

\begin{figure}[]
\begin{center}
\includegraphics[width=0.47\textwidth,clip,angle=0]{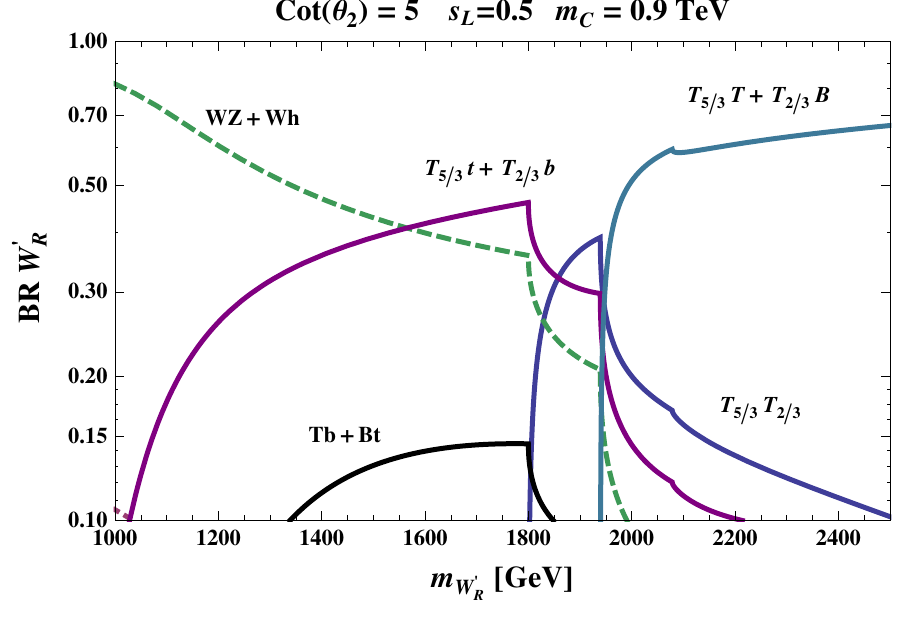}
\includegraphics[width=0.47\textwidth,clip,angle=0]{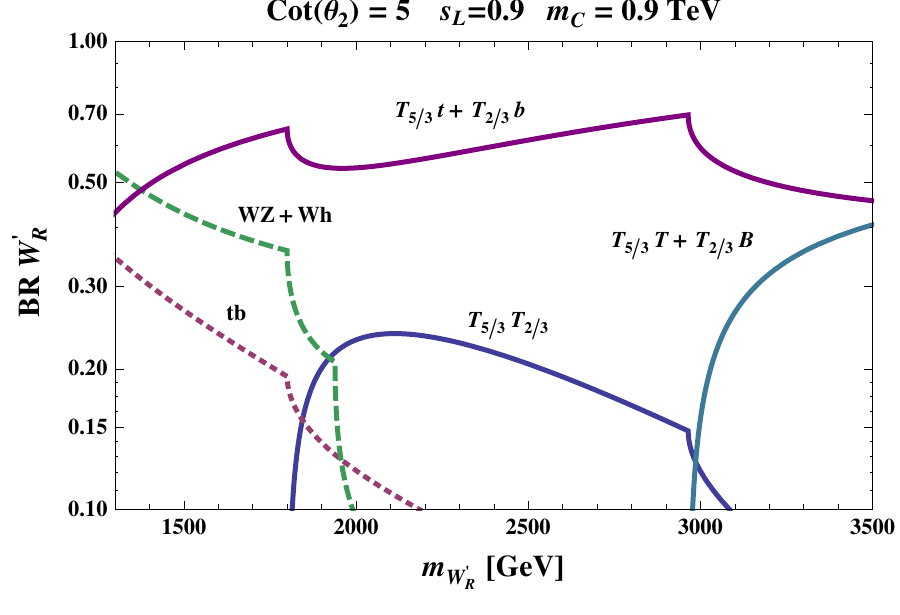}
\caption[]{
\label{fig:BRWR}
\small
$W^{'}_R$ decay branching fractions for $s_L=0.5$ (left panel) and $s_L=0.9$ (right panel). We set $\cot\theta_2=5$ and $m_C=0.9$ TeV.
}
\end{center}
\end{figure}

The $ W^{'}_R$ decay rates can be obtained from (\ref{eq:decay-mix}) by interchanging $s_{\xi} \leftrightarrow c_{\xi}$. The $ W^{'}_R$ decay branching fractions are shown in Fig. \ref{fig:BRWR} for $s_L=0.5$ (left panel) and $s_L=0.9$ (right panel) and for $\cot\theta_2=5$. Again, we have fixed $m_C=0.9$ TeV.\\

The dominant decay modes for large top degrees of compositeness are the channels of $ W^{'}_R$ decay into a custodian particle plus a SM quark, $T_{5/3}t, T_{2/3}b$. These decays remain relevant also for smaller $s_L$ values in the intermediate mass region. Promising signatures for the $ W^{'}_R$ discovery at the LHC are thus $T_{5/3}t \to Wtt$ and $T_{2/3}b \to Ztb, htb$.
For lighter $ W^{'}_R$, the $WZ, Wh$ channels could be also promising signatures.


\end{document}